\documentclass[twocolumn,showpacs,preprintnumbers,amsmath,amssymb,apsrev]{revtex4}

\usepackage{graphicx}
\usepackage{dcolumn}
\usepackage{bm}
\usepackage{natbib}

\begin{document}

\title{Structure and coarsening at the surface of a dry three-dimensional aqueous foam}
\author{A. E. Roth$^1$, B. G. Chen$^{1,2}$, and D. J. Durian$^1$}
\affiliation{
	$^1$Department of Physics \& Astronomy, University of Pennsylvania, Philadelphia, PA 19104-6396, USA \\
	$^2$Lorentz Institute, P.O. Box 9506, NL-2300 RA Leiden, The Netherlands
}

\date{\today}

\begin{abstract}
We utilize total-internal reflection to isolate the two-dimensional `surface foam' formed at the planar boundary of a three-dimensional sample.  The resulting images of surface Plateau borders are consistent with Plateau's laws for a truly two-dimensional foam.  Samples are allowed to coarsen into a self-similar scaling state where statistical distributions are independent of time, except for an overall scale factor.  There we find that statistical measures of side number distributions, size-topology correlations, and bubble shapes, are all very similar to those for two-dimensional foams.  However the size number distribution is slightly broader and the shapes are slightly more elongated.  A more obvious difference is that T2 processes now include the creation of surface bubbles, due to rearrangement in the bulk.  And von~Neumann's law is dramatically violated for individual bubbles.  But nevertheless, our most striking finding is that von~Neumann's law appears to holds on average.  Namely the average rate of area change for surface bubbles appears to be proportional to the number of sides minus six, but with individual bubbles showing a distribution of deviations from this average behavior.
\end{abstract}


\maketitle

\section{Introduction}

The structure and coarsening of three-dimensional foams is a topic that has long been of interest \cite{WeaireHutzlerBook}.  However, characterizing the microstructure in the bulk is difficult, and generally involves the use of sophisticated techniques beyond direct visual observation \cite{Matzke}.  This includes magnetic resonance imaging \cite{Gonatusetal95}, optical tomography \cite{Adler98, Fetterman00}, and x-ray tomography \cite{GlazierGraner05, GlazierGraner10, Meagher11}.  Foam microstructure is further difficult to measure because it changes with time.  Even if drainage and film ruptures are prevented, pressure differences between bubbles result in gas diffusion across films, such that some bubbles grow and others shrink.  This coarsening process is not limited to foams and is observed in other systems as well \cite{GlazierWeaire92, Stavans93}.  There have been measurements of coarsening in three-dimensional foams using light scattering \cite{DurianWeitzPine91a, DurianWeitzPine91b}, but such measurements involve an average of the system as a whole, and cannot probe individual bubble level.  For the case of ideal dry three-dimensional foams, there is an exact theoretical solution for the growth rate of an individual bubble with $n$ edges, which takes the form \cite{MacPhersonSrolovitz2007}
\begin{equation}
\frac{{\rm d}V}{{\rm d}t} = K \left(\sum_{i=1}^{n} e_{i} - 6L \right),
\label{mps}
\end{equation}
where $e_{i}$ is the length of edge $i$, and $L$ is a quantity called the `mean length' that depends on the size and shape of the bubble.  The constant, $K$, is proportional to film surface tension, the solubility and diffusivity of the gas, and the reciprocal of film thickness.  NMR \cite{Gonatusetal95} and tomography \cite{Adler98, GlazierGraner07} have been used to probe coarsening, but contact has not yet been made with Eq.~(\ref{mps}).

Much research on coarsening has been done for two-dimensional foams, where there are no difficulties in imaging the full microstructure.  This includes direct measurements of bubbles compressed between parallel plates \cite{GlazierGrossStavans87, GlazierStavans89, StavansGlazier89, Stavans90, Stavans93sf, Icaza94, RothDurian13}, soap froths with different boundary conditions \cite{StavansKrichevsky92, RosaFortes99, RosaFortes02}, and experiments on lipid monolayers \cite{SteinMoore90, Bergeetal90}.  There have also been simulations of two-dimensional foams \cite{KermodeWeaire90, GlazierAndersonGrest90, HerdtleAref92, Segeletal93, NeuSch97, Rutenberg05, Graneretalarxiv}.  These foams are simpler not only because of greater ease of measurement, but also due to simpler geometric considerations.  In particular, the coarsening rate of an individual bubble depends only on its number $n$ of sides according to the celebrated von~Neumann's law \cite{VonNeumann}:
\begin{equation}
\frac{{\rm d}A}{{\rm d}t} = K (n - 6).
\label{vn}
\end{equation}
The proportionality constant $K$ is not the same as in Eq.~(\ref{mps}), but it has a similar dependence on physicochemical properties and also has units of area per time.

The surface of a three-dimensional foam is where two and three dimensions meet.  When a three-dimensional foam is in contact with a flat two-dimensional surface, the films meet the surface at right angles, and the resulting network of surface Plateau borders meet at 120$^\circ$ at three-fold vertices \cite{Mancini05a, Mancini05b}.  Euler's law thus implies that the average number of sides, for a large sample, should be six.  Thus the `surface foam' looks very much like a two-dimensional foam, and the natural question we investigate here is the ways in which they are quantitatively different.

Surface foams have been of previous interest, mostly as a way of connecting to the properties of the larger three-dimensional foam.   This is especially important for situations where it is not feasible to measure the full three dimensional structure.  For example, foams with even a small nonzero liquid fraction are opaque so only surface bubbles can be imaged.  Three-dimensional imaging tends to be slow, so it's also useful to consider surface bubbles for foams under shear.  Prior work includes experiments on the radial distribution of very wet foams with nearly spherical bubbles \cite{ClarkBlackman48, ChangSchoenGrove56, deVries57}, experiments on the surface of continuously bubbled foams \cite{JashnaniLemlich74, FeitosaDurian06}, and work on the effect of liquid fraction \cite{Jameson99}.  There has also been theoretical work on the conversion of surface measurements to bulk measurements \cite{deVries57, deVries72, ChengLemlich}, as well as Surface Evolver simulation comparisons of surface and bulk properties for very dry foams \cite{WangNeethling06, WangNeethling09}.  And there has been theoretical work comparing the structure of two-dimensional coarsening systems with cross sections of three dimensional systems \cite{LazarSrolovitz}.

If left to coarsen, both two- and three-dimensional foams are believed to reach a self-similar scaling state where, apart from an overall scale factor, statistical distributions of bubble size, shape, and topology are independent of time.  Therefore, we expect that surface foams will also reach a scaling state -- though with different statistics from a truly two-dimensional foam.  In part this is because boundary bubbles coarsen at a different rate from bulk bubbles.  In two dimensions, the von~Neumann argument can be extended to boundary bubbles by summing the diffusive flux across interior films and using the fact that films terminate at the boundary at right angles~\cite{HerdtleAref92, RosaFortes02a}.  For the case of a flat boundary, the result is
\begin{equation}
\frac{{\rm d}A}{{\rm d}t} = K (n - 5).
\label{vnbound}
\end{equation}
Thus coarsening still only depends on the number of sides, but now 5 sided bubbles are stationary, and 6 sided bubbles grow.  We have similarly calculated the growth rate for a three-dimensional bubble in contact with a flat boundary, using the same geometric method as MacPherson and Srolovitz \cite{MacPhersonSrolovitz2007}.  For a planar boundary, we find
\begin{equation}
\frac{{\rm d}V}{{\rm d}t} = K \left[\sum_{interior} e_{i} + \left(\frac{3}{2} \sum_{boundary} e_{i} \right) - 6L \right]
\label{mpsbound}
\end{equation}
This is similar to Eq.~(\ref{mps}), but now the contribution from interior edges is different from boundary edges due to weighting by their respective dihedral external angles of $\pi/3$ and $\pi/2$.  Note that Eq.~(\ref{mpsbound}) describes the growth rate of bubble volume, and not the growth of the face area in contact with the boundary.  

Bubble dynamics also include topological changes of the foam.  These topological changes come in two types: T1 rearrangement processes, which do not change the total number of bubbles, and T2 processes, which involve the creation or disappearance of a bubble.  In both surface foams and two-dimensional foams, T1 processes and the subset of T2 processes that involve the disappearance of a bubble will look the same.  However, in two-dimensional foams, bubble creation by coarsening is not possible.  In surface foams, by contrast, the movement of a bulk bubble to the surface will result in the apparent creation of a bubble.

In general it is difficult to photograph just the surface of a three-dimensional dry foam because both surface and interior Plateau borders come into focus and cannot be readily distinguished.  To overcome this effect and image only the surface Plateau borders, we have developed an imaging technique based on total internal reflection.  We thus obtain clear images of `surface foams', where the  `bubbles' consist of the co-planar exterior faces of three-dimensional boundary bubbles in a bulk foam.  We then use standard digital analysis methods to extract the properties of individual bubbles, just as per ordinary two-dimensional foams.  We also track the growth rate of individual bubbles over time.

\section{Materials and Methods}

\begin{figure}
\includegraphics[width=3.00in]{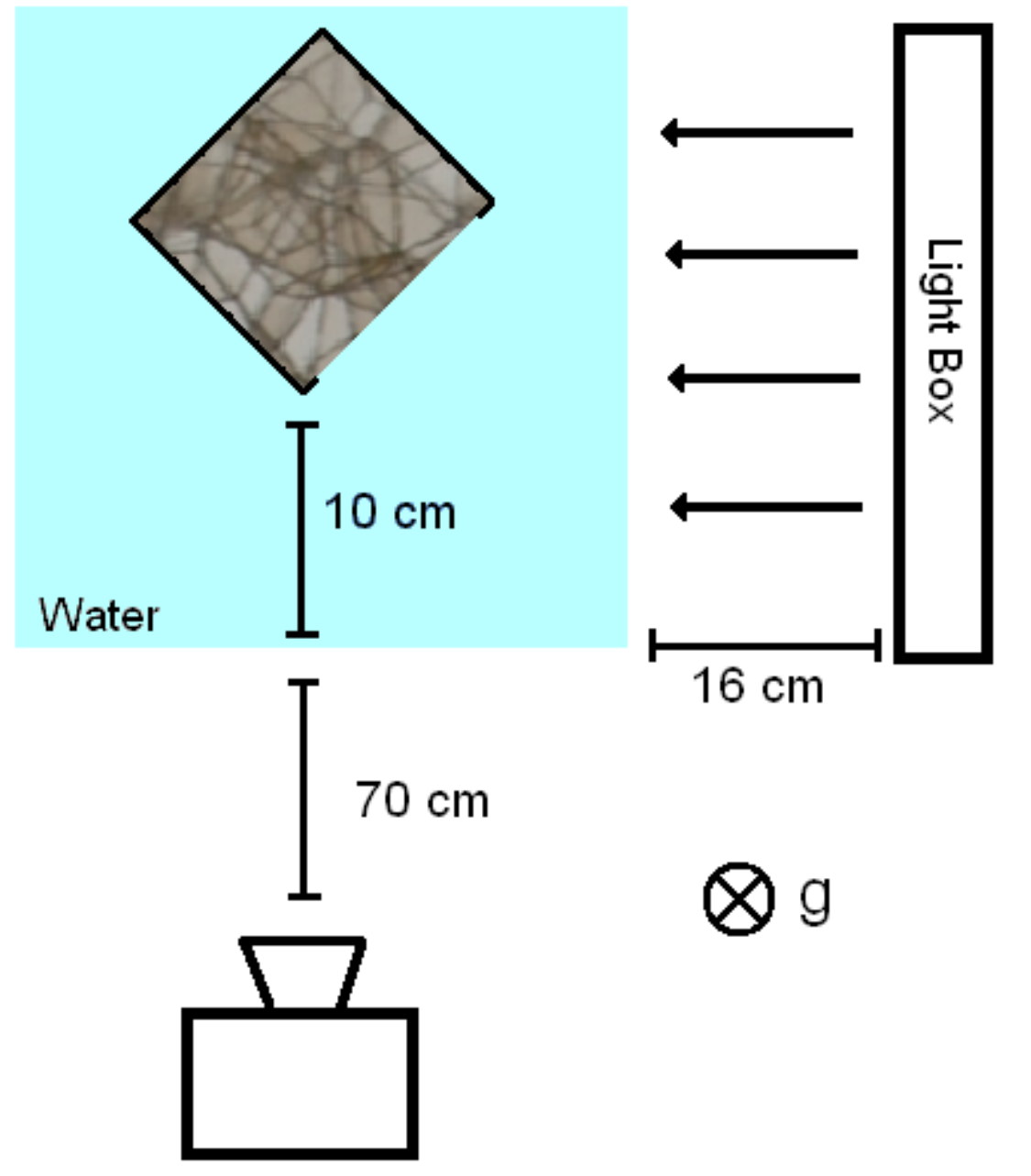}
\caption{Schematic diagram of a top-down view of the imaging setup.  The square bottle in the center is filled with foam and submerged in a tank of water.  On the right of the diagram is a lightbox that provides constant uniform illumination.  At the bottom of the diagram is a camera to image the surface of interest.  Not to scale.}
\label{diagramsurface}
\end{figure}

Our apparatus is depicted in Fig.~\ref{diagramsurface}.  The foam sample is inside a sealed plastic bottle with a square cross section of $9~{\rm cm} \times 9~{\rm cm}$ and a height of 14~cm.  All areas of the bottle are masked with electrical tape except for a single flat surface of interest.  The sample bottle is completely submerged in a square tank of water.  A Vista Point A lightbox is placed to the side of the tank to provide steady, uniform illumination.  A Nikon~D80 camera with an AF-S Nikkor 55-200~mm 1.4-5.6G ED lens is pointed at the face of the tank $90^{\circ}$ from the lightbox.  The bottle is placed so that the surface of interest is at an angle relative to the lightbox and the camera.

To create foam, 275~mL of a solution consisting of 75\% deionized water, 20\% glycerin, and 5\% Dawn ultra concentrated dish detergent is sealed into the bottle and vigorously shaken.  This gives a polydisperse distribution of bubbles with an average diameter less than 0.5~mm.   Data collection begins after two hours, when the foam is very dry due to drainage and the average bubble diameter is more than ten times larger due to coarsening.   The number of bubbles in the sample at production, and at the commencement of imaging, are of order $10^7$ and $10^4$, respectively.  Film ruptures are never observed in the course of our experiments.

\begin{figure*}
\includegraphics[width=6.00in]{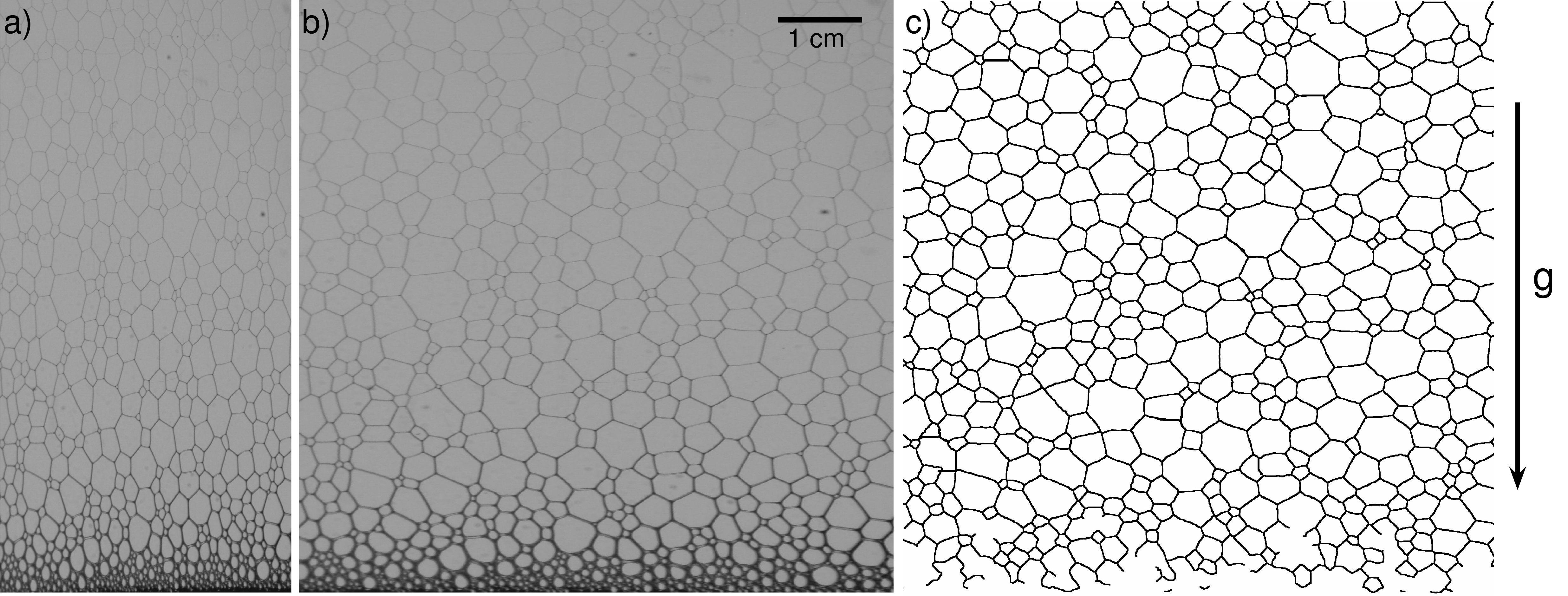}
\caption{(a) Example raw image of the a surface foam at half a day after production, along with (b) transformed image to correct for distortion and (c) skeletonized digital image suitable for analysis.}
\label{transform}
\end{figure*}

The apparatus allows the surface foam to be isolated, as follows.  The angle of the bottle is chosen so that if a light ray strikes a point on the surface of the bottle that has the interior of an air bubble on the other side, it will be totally internally reflected in the specular direction toward the camera.  But if the light ray strikes a surface Plateau border, it will be preferentially refracted into another direction. This gives raw images with bright cells and dark Plateau borders, as shown in an example in Fig.~\ref{transform}a.  Only surface Plateau borders are visible, as desired.

Note that the raw image in Fig.~\ref{transform}a is distorted, since the bottle is at an angle, and this must be corrected.  Fiducial marks are made at the corners to define a rectangular region with right angle corners.  Using the position of these marks, it is possible to transform the image to a direct perspective.  A sample image of the surface foam after this transformation is shown in Fig.~\ref{transform}b.  After transformation, thresholding and skeletonization are performed to identify the cells, separated by skeletonized borders.  The results of this image analysis can be seen in Fig.~\ref{transform}c.  From this we extract relevant quantities for the individual bubbles, such as area and number of sides, using standard procedures as described in Ref.~\cite{RothDurian13}.  We only consider bubbles that lie completely within a region of interest is higher than a couple centimeters above the drained liquid, where the average bubble size is independent of height as seen in Fig.~\ref{transform}.  Occasionally the skeletonization procedure removes a film, as can be seen Fig.~\ref{transform}c.  Such mistakes are detected automatically during bubble identification, and are fixed by hand.  For noisier images, it might be advantageous to use the reconstruction method of Ref.~\cite{VCG03} since it is immune to tracking errors.

Before proceeding, we note that the average size of the surface bubbles in the coarsened foam of Fig.~\ref{transform} are independent of height above a couple centimeters from the bottom.  We restrict attention to this uniform region.  There, the coarsening rate must be independent of height, else there would be a noticeable vertical gradient in average bubble size.  Consequently, the film thickness must also be nearly constant, with essentially negligible thinning due to gravity.  For our system, the effective interface potential that controls film thickness must have a minimum that is very steep compared to gravity.  Indeed, this same behavior was observed directly in Fig.~4 of Ref.~\cite{RothDurian13}, where the rate of area change was plotted versus height for a few hundred quasi-2d bubbles made with the same surfactant system as here.

\section{Bubble Distributions}

\subsection{Topology}

\begin{figure}
\includegraphics[width=3.00in]{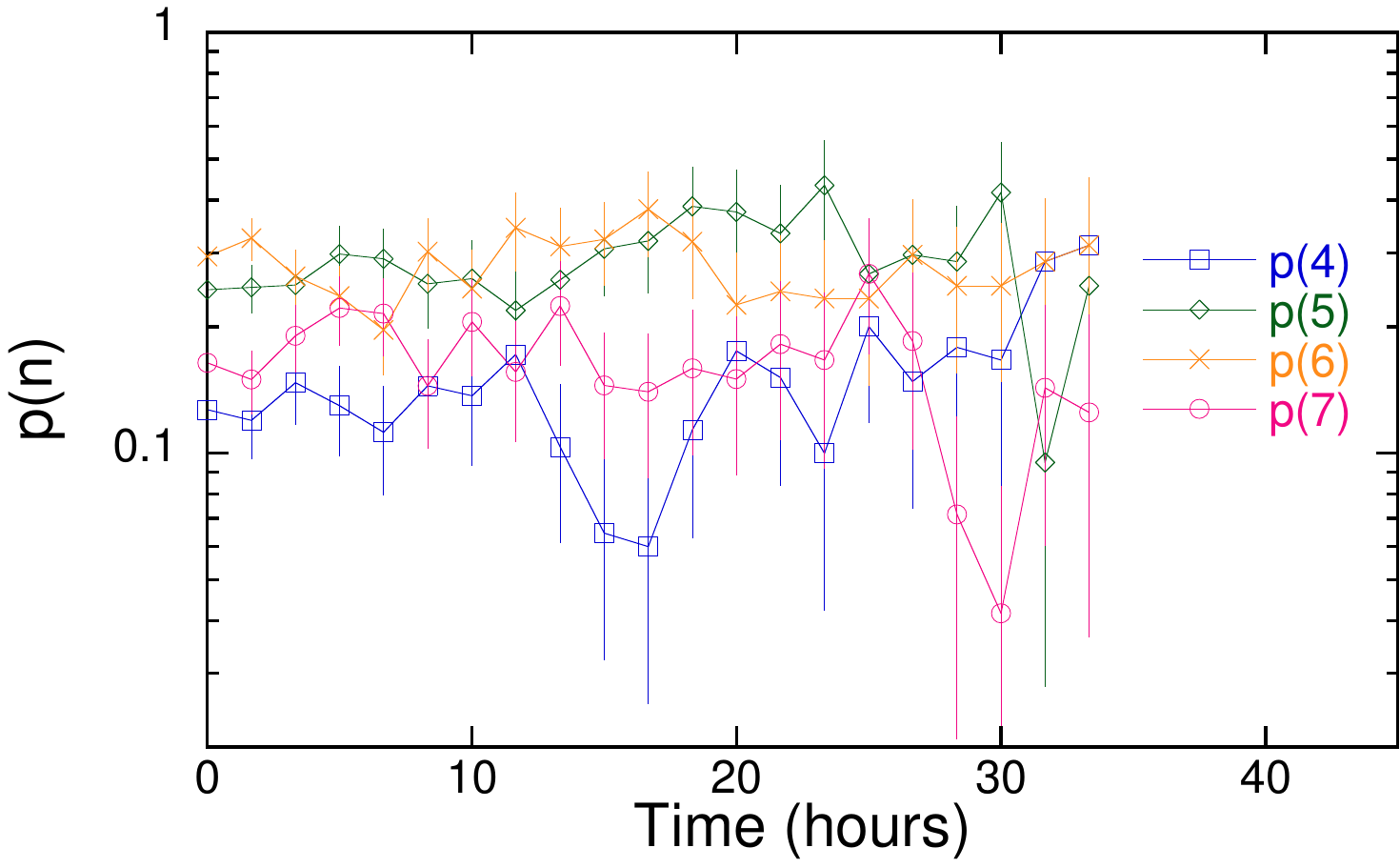}
\caption{(Color online) Side number distribution, $p(n)$, versus time for 4, 5, 6, and 7 sided bubbles for a single run.  Time zero for data collection is two hours after foam production, when the average diameter is ten times greater than at initial foam production.}
\label{pnvtsurface}
\end{figure}

The first relevant quantity to consider for a foam is the distribution of the number of sides, $p(n)$.  This is the probability that a randomly selected bubble from the foam will have $n$ sides.  Probabilities for $n\in\{4,5,6,7\}$ are shown versus time in Fig.~\ref{pnvtsurface} for one foam sample.  The data are noisy, but there does not appear to be any systematic change over the 30~hour period starting two hours after production.  In other words, fits of $p(n)$ to a linear function of time all give a slope that is within error of zero.   This is consistent with the foam being in a self-similar scaling state, where distributions do not change shape with time.  This is expected, since the initial foam is polydisperse and the average diameter increased by a factor of ten prior to data collection.  The time to reach the scaling state is faster for polydisperse samples, but only a factor of ten in diameter growth is required even for monodisperse samples \cite{FeitosaDurian06}.

\begin{figure}
\includegraphics[width=3.00in]{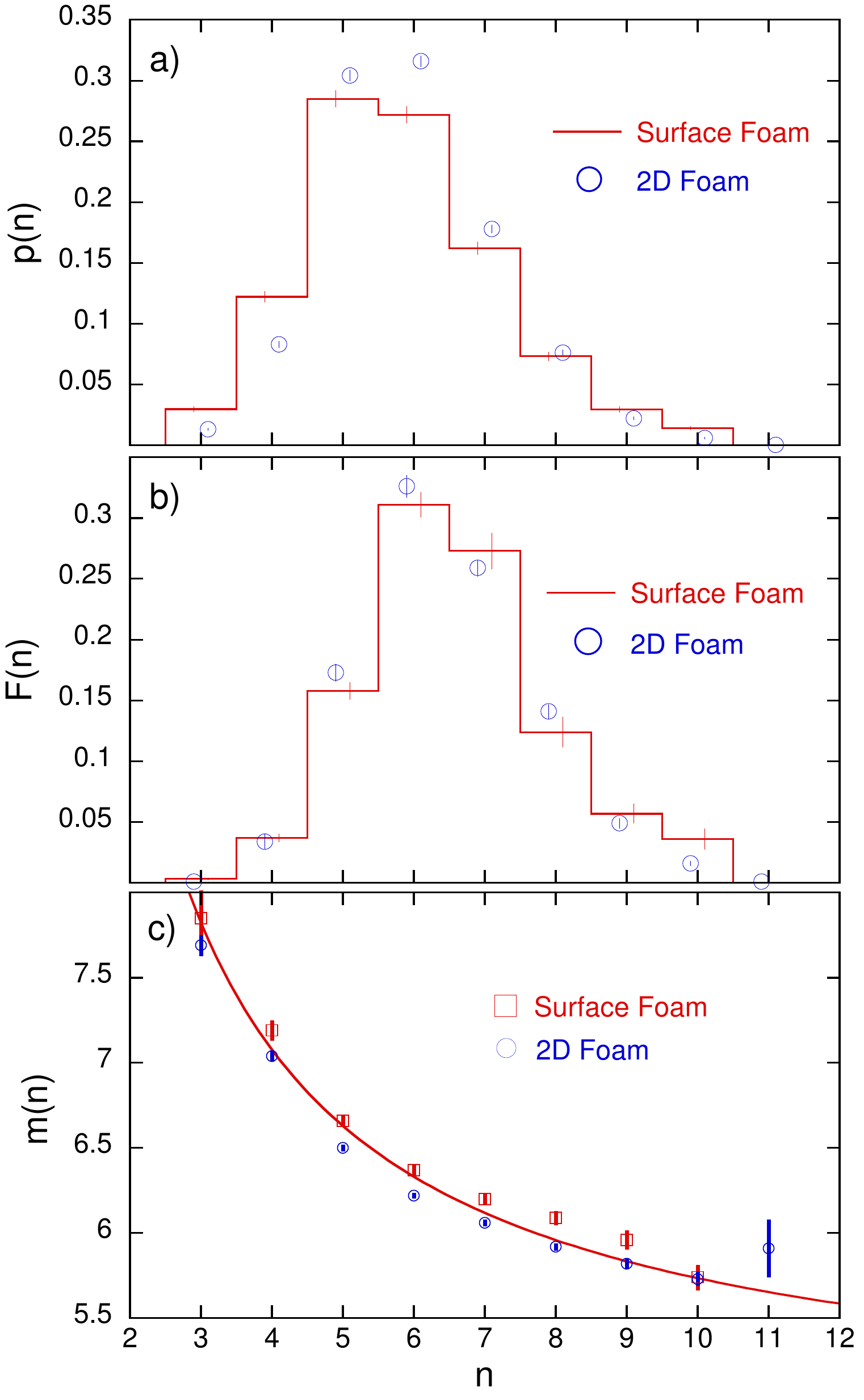}
\caption{(a)  Side number distribution, $p(n)$, averaged over all times for all runs.  Bubbles with $n < 3$ or $n > 10$ were not observed. Total number of bubbles is 5966.  Average number of sides is $\langle n \rangle = 5.83 \pm 0.02$.  Data for the two-dimensional foam is taken from Ref.~\cite{RothDurian13}.  (b)  Area weighted side number distribution, $F(n)$, averaged over all time for all runs.  Area weighted average number of sides is $\langle \langle n \rangle \rangle = 6.6 \pm 0.2$  Data for the two-dimensional foam is taken from Ref.~\cite{RothDurian13}.  (c)  $m(n)$ is the average number of sides of the neighbors of an $n$-sided bubble.  The data is averaged over all times for all runs.  The solid curve is the Aboav-Weaire law, $m(n) = (6 - a) + (6a + \mu_{2})/n$, where $\mu_{2}$ is the variance, $\langle (n-\langle n \rangle)^{2} \rangle$, of the side distribution (for our system $\mu_{2} = 1.99 \pm 0.04$) and $a$ is the only fitting parameter, which we measure to be $a = 1.16 \pm 0.07$.  Data for the two-dimensional foam is taken from Ref.~\cite{RothDurian13}.}
\label{pnfnmn}
\end{figure}

The distribution of number of sides is also found to be the same, to within statistical error, for four different runs.  This allows us to average the distribution, $p(n)$, over all times and for all runs, which comprise a total of 5966 different bubbles.  The overall side number distribution is shown in Fig.~\ref{pnfnmn}a.  Also shown for comparison is the side number distribution for an ordinary two-dimensional foam.  We see that there is a difference in the distributions.  Even though the surface foam obeys Plateau's laws and looks in that way like a two-dimensional foam, the different dynamics lead to a different scaling state.  We see that the distribution for the surface foam is broader, with fewer five and six sided bubbles and more four sided bubbles.  This means that the surface foam has a higher variance, $\mu_{2} = \langle (n-\langle n \rangle)^{2} \rangle$, which we measure to be $\mu_{2} = 1.99 \pm 0.04$, as compared to the value measured for the two-dimensional foam, $\mu_{2} = 1.56 \pm 0.02$ \cite{RothDurian13}.  The average number of sides of the surface foam is $\langle n \rangle = \sum np(n) = 5.83\pm0.02$, which is less than the required value of 6 only because the sample is of finite size.

A related distribution that is less well-known is the area-weighted side number distribution, $F(n)$.  This is defined in Ref.~\cite{RothDurian13} as the probability that a randomly selected point within the foam falls inside an $n$-sided bubble.  When calculating the growth rate of the average area for a two-dimensional foam, the rate depends on the quantity $\sum_{n} nF(n)$.  As with $p(n)$, the distribution $F(n)$ does not vary with time, and so we can average over all times for all runs. The results are shown in Fig.~\ref{pnfnmn}b.  Also shown for comparison is the distribution of $F(n)$ for an ordinary two-dimensional foam.  The area-weighted average number of sides is $\langle\langle n \rangle\rangle = \sum nF(n)=6.6\pm0.2$, which is slightly larger but within error of the result for two-dimensional foams~\cite{RothDurian13}.

The average number $m(n)$ of sides of an $n$-sided bubble is another topological quantity of interest.  As with the side number distribution, this quantity does not change over time or for the different runs.  This allows us to average over all $n$-sided bubbles for all times and for all runs.  The results are shown in Fig.~\ref{pnfnmn}c.  The expected form of this relationship, known as the Aboav-Weaire law, is  $m(n) = (6-a) + (6a + \mu_{2})/n$, where $\mu_{2}$ is the variance, and $a$ is the only fitting parameter.  We find $a = 1.16 \pm 0.07$, which is within error of measurements for ordinary two-dimensional foams \cite{ChiuReview, WeaireHutzlerBook, RothDurian13}.

\subsection{Size}

\begin{figure}
\includegraphics[width=3.00in]{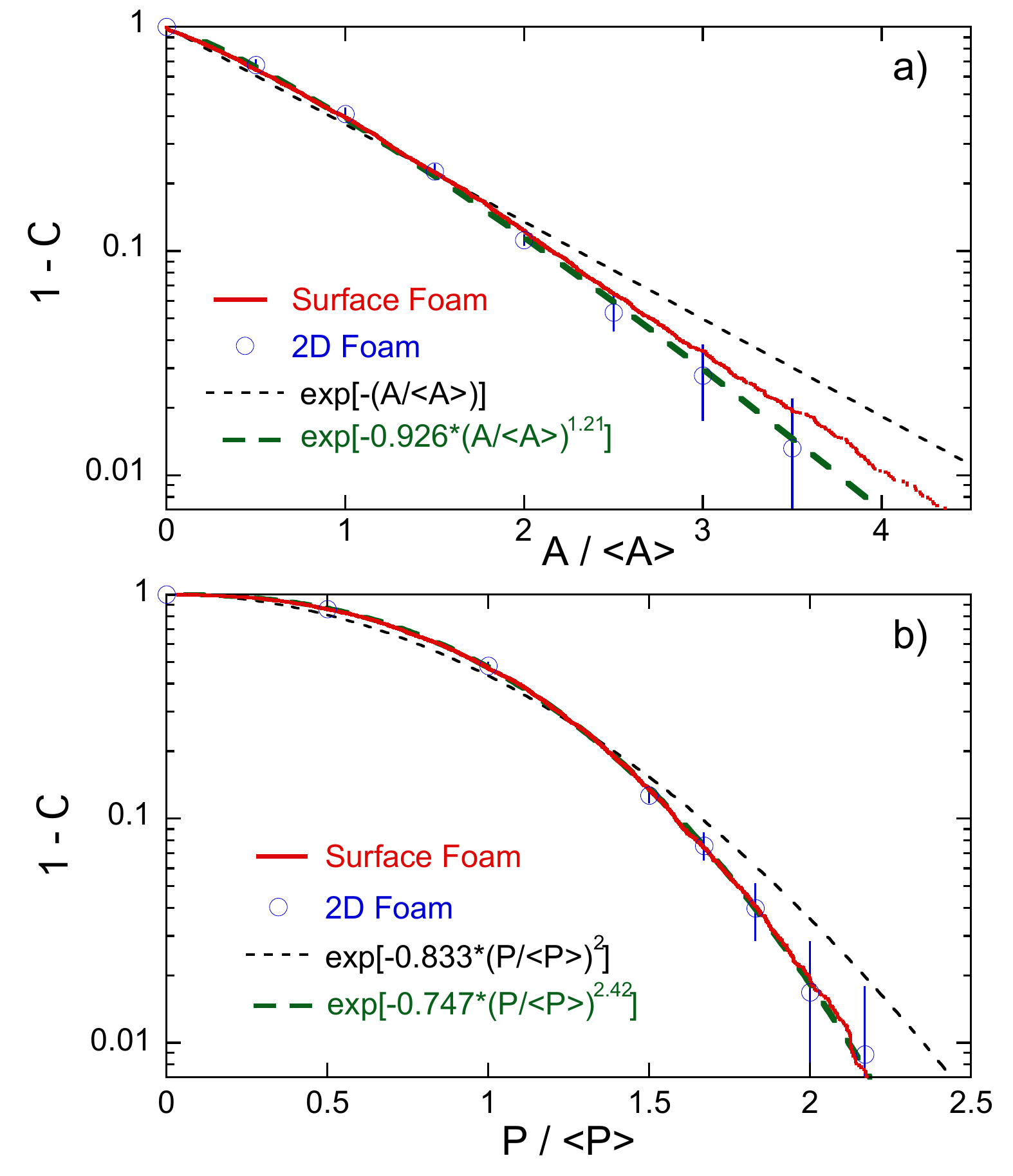}
\caption{One minus the cumulative distribution of (a) bubble areas and (b) bubble perimeters, averaged over all times for all runs.  The dotted black line is an exponential shown for comparison.  The green dashed curve is fit to a compressed exponential.  The compressed exponential in (b) corresponds to the form in part a), assuming that $A \propto P^{2}$ with the same proportionality constant for all bubbles.  Data for the two-dimensional foam, and the compressed exponential curves, are taken from Ref.~\cite{RothDurian13}.}
\label{areadistsurface}
\end{figure}

The distribution of bubble areas is one natural measure of bubble size.  Although the average area of the bubbles increases with time, if we divide out the average area then the distribution of the resulting normalized area does not change with time, and is found to be the same for all runs, to within statistical uncertainty.  This is consistent with the foam being in a scaling state.  Therefore it is possible to average the normalized area distribution for all times and for all runs.  The cumulative distribution of bubble areas is shown in Fig.~\ref{areadistsurface}a.  The curve corresponding to an exponential distribution is shown for comparison as a dotted line.  Our data falls below the exponential curve for large $A / \langle A \rangle$ and is better fit by a compressed exponential, shown as a dotted line.  We find that the cumulative area distribution for the surface foam is very similar to the distribution for a two-dimensional foam.  While it falls above the two-dimensional data for large $A/\langle A \rangle$, this deviation is within the error bars.

As with the area, we measure the perimeter of each bubble.  We average the normalized perimeter across all times and runs.  The cumulative distribution of perimeters can be seen in Fig.~\ref{areadistsurface}b.  The normalized perimeter falls below the exponential curve, and is well fit by a compressed exponential.  The compressed exponential shown for the perimeter distribution corresponds to the compressed exponential for the area distribution, assuming that $A \propto P^{2}$ with the same proportionality constant for all bubbles.  This form is the same as Eqs.~(7-8) in Ref.~\cite{RothDurian13}.  We see that the perimeter distribution for the surface foam falls on top of the distribution for a two-dimensional foam and does not show the deviation for large bubbles seen in the area distribution.

\subsection{Size-Topology}

We have characterized the distribution of the number of sides and the area distribution, but it is also useful to look at quantities that depend on both these measurements in different ways.  One example is the average area of an $n$-sided bubble.  This is a relationship that has been of interest in the past \cite{ChiuReview}.  The first empirical measurements were made by Lewis for epithelial cucumber cells, who found a linear relationship of the form
\begin{equation}
\frac{\langle A_{n} \rangle}{\langle A \rangle} = 1 + \lambda(n-6)
\label{lewislaw}
\end{equation}
where $\lambda$ is a parameter of the system \cite{Lewis1928, Lewis1930}.  It can be shown that if $\langle A_{n} \rangle / \langle A \rangle$ is linear in $n$ then this relationship must hold, but additional restraints are required to prove that this relationship must be linear \cite{RivierLissowski82}.  A related measurement that is also of interest is the relationship between the average perimeter of an $n$-sided bubble and $n$.  This analogous relationship is called Desch's law or Feltham's law, and is of the same form as Lewis' law with the area replaced by perimeter.  Specifically, it has the form
\begin{equation}
\frac{\langle P_{n} \rangle}{\langle P \rangle} = 1 + \nu(n-6)
\label{deschlaw}
\end{equation}
where $\nu$ is a parameter of the system.  It has been shown that if the average energy of a cell is proportional to its perimeter, then the entropy is maximized if Desch's law is satisfied \cite{Rivier85}.  These laws continue to be of interest \cite{SzetoTam95, Saraivaetal09, Graneretal11}.

\begin{figure}
\includegraphics[width=3.00in]{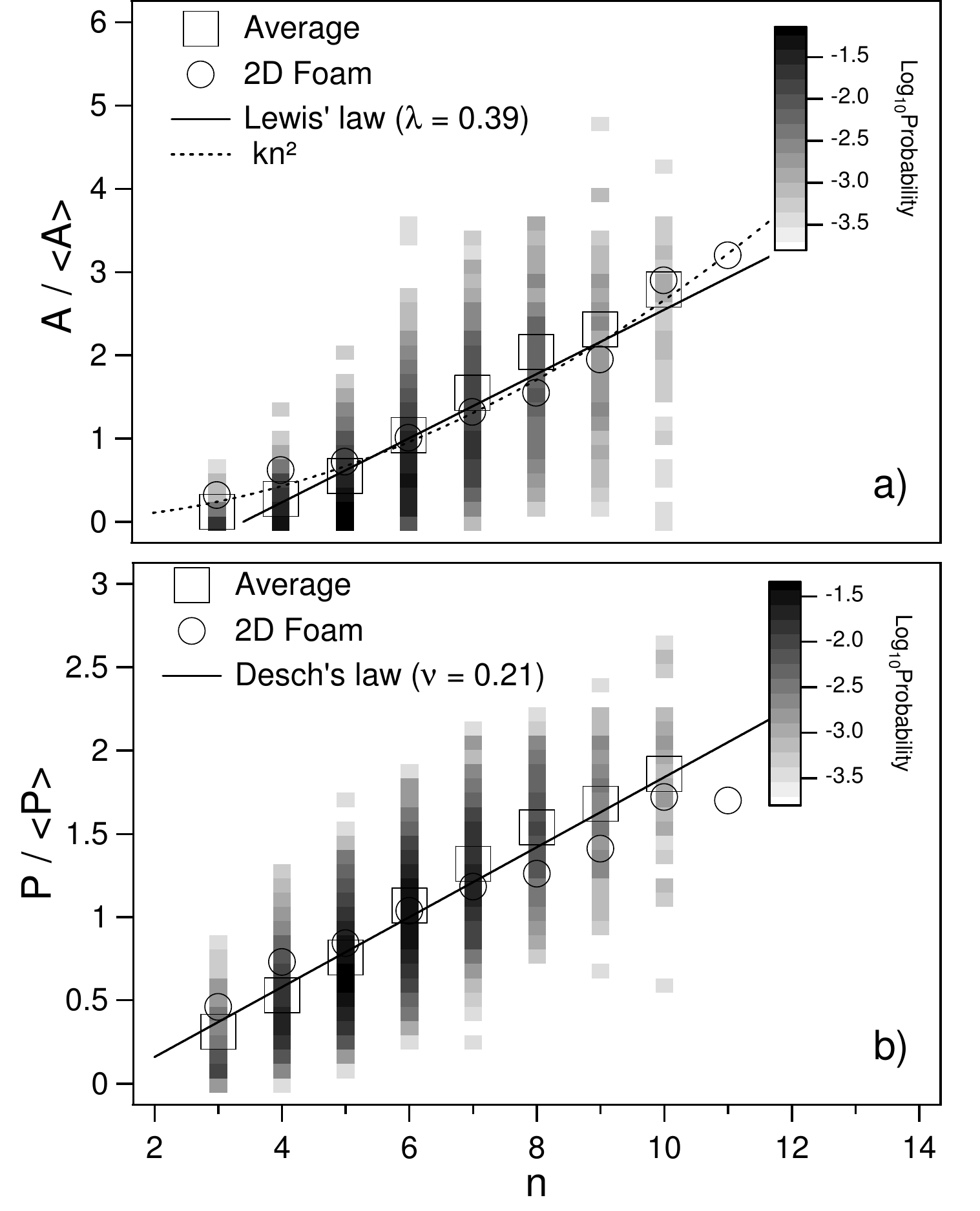}
\caption{(a)  Normalized area versus side number for all bubbles.  Grayscale shows the probability of finding a bubble with that side number and that area. Squares are the average normalized area for a given $n$, $\langle A_{n} \rangle / \langle A \rangle$.  The solid line is a fit to Lewis' law, $\langle A_{n} \rangle / \langle A \rangle = n \lambda + (1 - 6 \lambda)$, with fitted value $\lambda = 0.39 \pm 0.03$.  The dotted line is a fit to the proportionality $\langle A_{n} \rangle / \langle A \rangle = kn^{2}$.  (b)  Normalized perimeter versus side number for all bubbles.  Grayscale shows the probability of finding a bubble with that side number and that perimeter.  Squares are the average normalized perimeter for a given $n$, $\langle P_{n} \rangle / \langle P \rangle$.  Solid line is a fit to Desch's law, $\langle P_{n} \rangle / \langle P \rangle = n \nu + (1 - 6 \nu)$, with fitted value $\nu = 0.21 \pm 0.01$.  In both parts, data for the two-dimensional foam is taken from Ref.~\cite{RothDurian13}.}
\label{lewisdeschsurface}
\end{figure}

We measured $A / \langle A \rangle$ for all bubbles and the results, for all times and all runs, are shown versus side number in Fig.~\ref{lewisdeschsurface}a.  The grayscale corresponds to the probability of finding a bubble with that number of sides and that normalized area for each point.  The average, $\langle A_{n} \rangle / \langle A \rangle$, is shown as squares.  Note that the distribution of normalized areas around the average for a given $n$ is not symmetric and, especially for small $n$, is peaked near zero.  A fit to Lewis' law is shown as a solid line, and does not match the data closely.  This demonstrates that Lewis' law is not an appropriate fit for our data, which is fit better by a generic quadratic form, shown on the plot as a dotted line.  This result is in accordance with some simulations and experiments on ordinary two-dimensional foams \cite{SzetoTam95, Graneretal11, RothDurian13}.  We see that the values for the surface foam are not significantly different from the two-dimensional foam, although it looks as though the two-dimensional data may have slightly more curvature.  Both cases clearly deviate from Lewis' law.  This deviation from Lewis' law is consistent with our result that the area distribution deviates from an exponential \cite{Lambert2012}.

We similarly measured the normalized perimeters, $P / \langle P \rangle$, for all bubbles, and the results for all times and runs are shown in Fig.~\ref{lewisdeschsurface}b.  As in the plot for Lewis' law, the grayscale corresponds to the probability of finding a bubble with a given number of sides, $n$, and a given normalized perimeter.  The average for each $n$, $\langle P_{n} \rangle / \langle P \rangle$, is shown as squares on the plot. Note that the distributions about the average are much more symmetrical than in the area case shown in Fig.~\ref{lewisdeschsurface}a.  The fit to Desch's law, shown as a solid line on the plot, is a good fit to the data.  Desch's law is a better fit to our data than Lewis's law.  We see that the surface foam data is clearly linear, as is the data for the two-dimensional foam  However, the slopes of the lines in the two cases are different.  The value of $\nu$ measured for the surface foam in the fit to Eq.~(\ref{deschlaw}) is $\nu = 0.21 \pm 0.01$.  This value is in the same general range as measurements made for ordinary two-dimensional foams \cite{SzetoTam95, Graneretal11, RothDurian13}.

\subsection{Shapes}

\begin{figure}
\includegraphics[width=3.00in]{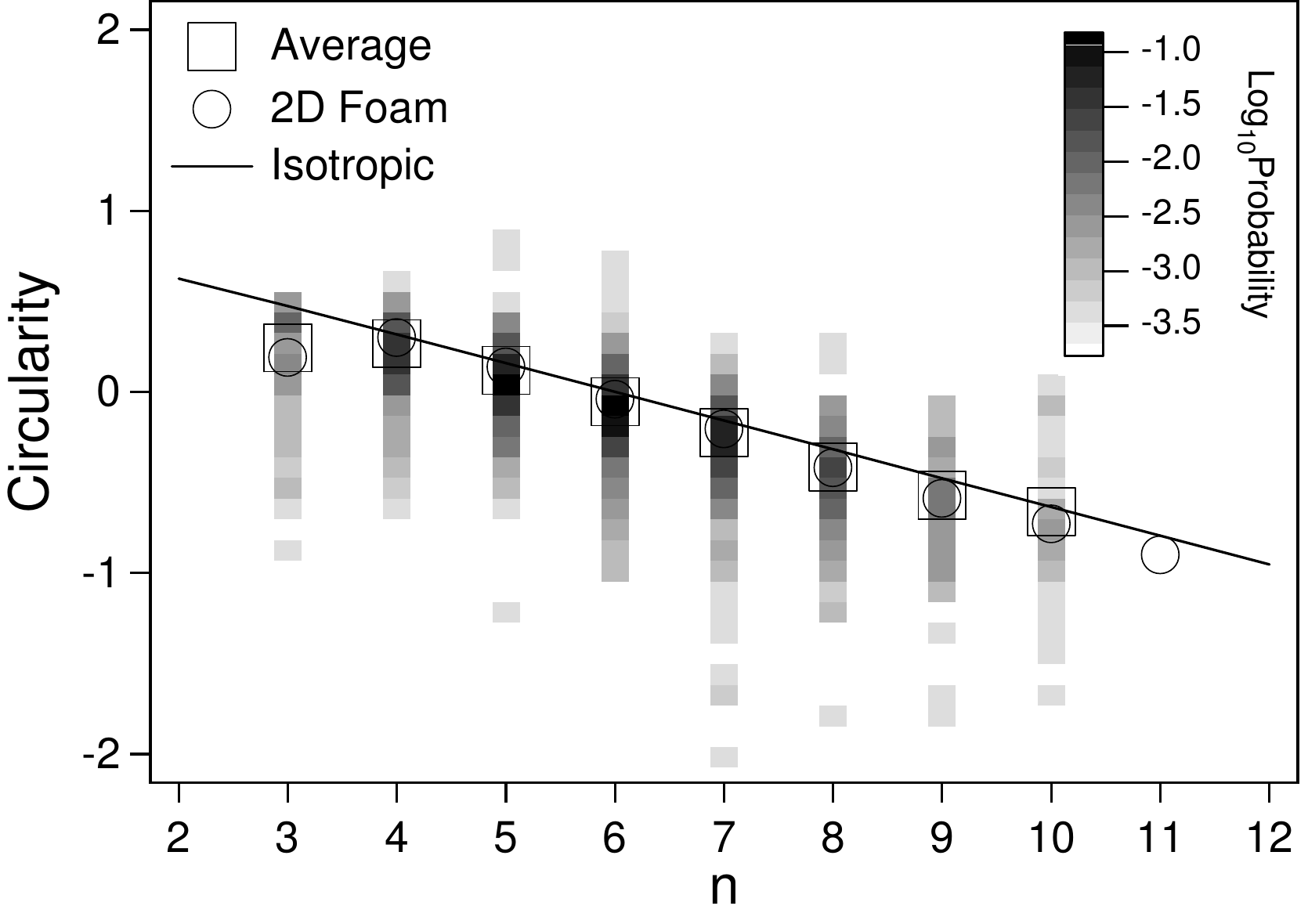}
\caption{Circularity, defined by Eq.~(\ref{circularity}), versus side number for all bubbles.  Grayscale shows the probability of finding a bubble with that side number and circularity, normalized so that the sum over $n$ and integral over circularity gives 1.  Squares are the average circularity for a given $n$.  The solid line shows the circularity for isotropic bubbles.  Average circularity data for two-dimensional foam, from Ref.~\cite{RothDurian13}, is also shown for comparison.}
\label{circularitysurface}
\end{figure}

\begin{figure}
\includegraphics[width=3.00in]{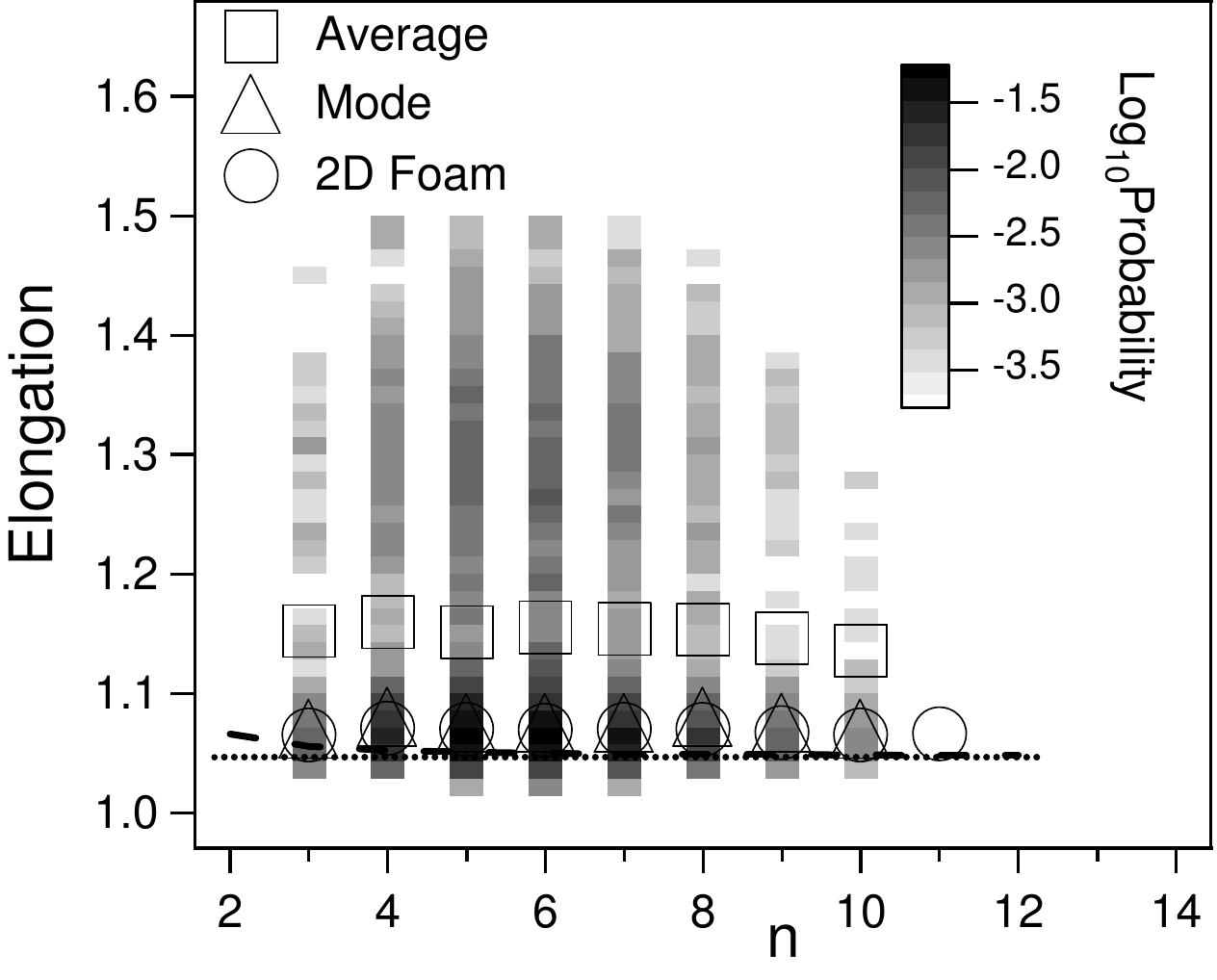}
\caption{Elongation, $E=P/\sqrt{4 \pi A}$ where $P$ is perimeter and $A$ is area, plotted versus side number.  Grayscale shows the probability of finding a bubble with that side number and that elongation, normalized so that the sum over $n$ and integral over elongation gives 1.  Squares are the average elongation for a given $n$.  The dashed curve shows the elongation for isotropic bubbles with $n$ sides.  The dotted line is the limit of the elongation of an isotropic $n$-sided bubble as $n$ goes to infinity.  The triangles are the mode of the circularity for a given $n$.  Average elongation data for two-dimensional foam, from Ref.~\cite{RothDurian13}, is also shown for comparison.}
\label{elongationsurface}
\end{figure}

There are many ways to characterize the shape of a bubble.  Among the possible shape parameters, there are two that have special physical significance with regards to the coarsening process of dry two-dimensional foams \cite{RothDurian13} with nonzero liquid content.  The first is circularity, defined as
\begin{equation}
	C = \left( \frac{1}{n} \sum^{n}_{i} 1 / R_{i} \right) \sqrt{A/\pi},
\label{circularity}
\end{equation}
where $A$ is the area and $R_{i}$ is the radius of curvature for the $i^{th}$ side of an $n$-sided bubble.   This dimensionless number is 1 for a circle and 0 for any shape made up of straight line segments. The sign convention is such that $R_{i}$ is positive for the bubble on the high-pressure side of the film.  Note that the surface Plateau borders must be circular arcs for $C$ to be well defined, which in turn requires the curvature of the films perpendicular to the surface to be constant.  While the films must certainly meet the boundary at $\pi/2$, it is not obvious that the curvature conditions holds.  However, we find that all surface Plateau borders may be well-fit to circular arcs with no systematic deviation to within the accuracy of the data.  The second relevant shape parameter is elongation, defined as
\begin{equation} 
	E=P/\sqrt{4 \pi A} 
\label{elongation}
\end{equation}
where $P$ is the perimeter and $A$ is the area.  This dimensionless number is 1 for a circle, and a large elongation would correspond to a shape far from a circle.

The distribution of circularities for all times and for all runs is shown in Fig.~\ref{circularitysurface}.  The grayscale corresponds to the probability of finding a bubble with that circularity and that number of sides.  The average circularity for each $n$ is shown as squares.  For comparison, the circularity of an isotropic bubble is shown as a straight line.  An isotropic bubble is an $n$-sided bubble with all sides the same length and having the same curvature.  We see that the circularity for two-dimensional foams and surface foams is very similar.  Both are similar to the isotropic case, except for three sided bubbles, which fall below the line.

The distribution of elongations, for all times and for all runs, is shown in Fig.~\ref{elongationsurface}.  The grayscale corresponds to the probability of finding a bubble with that elongation and that number of sides.  The average elongation for each $n$ is shown as squares.  This average elongation does not depend on $n$.  For comparison, the elongation of an isotropic bubble is shown as a dashed line.  We see that, unlike the circularity, the values for isotropic bubbles are not close to the average value.  The overall shape of the distribution has a main peak, with a long tail and small secondary peak (note that the probabilities are shown on a logarithmic scale).  This long tail and secondary peak increases the average elongation, but the mode, indicated by circles on the plot, shows that the distribution is peaked near the value for an isotropic bubble.  Ordinary two-dimensional foams have a distribution of elongations that is much less broad.  There is no tail of highly elongated bubbles.  We see that this tail for surface bubbles causes the average value to be far from the peak of the distribution.  Although the average elongation for the surface foam and the two-dimensional foam are far apart, the average of the surface foam is close to the peak of the distribution for the surface foam.  This suggests that most surface bubbles have an elongation in a range similar to what we see for two-dimensional foams, but in surface foams there exists a tail of highly elongated bubbles not present in ordinary two-dimensional foams.  The fact that the elongation distribution is different for the surface foam and the two-dimensional foam is consistent with the fact that the area distribution for the surface foam deviates from the area distribution for the two-dimensional foam, but the perimeter distributions are the same in both cases.

\section{Bubble Dynamics}

\begin{figure}
\includegraphics[width=3.00in]{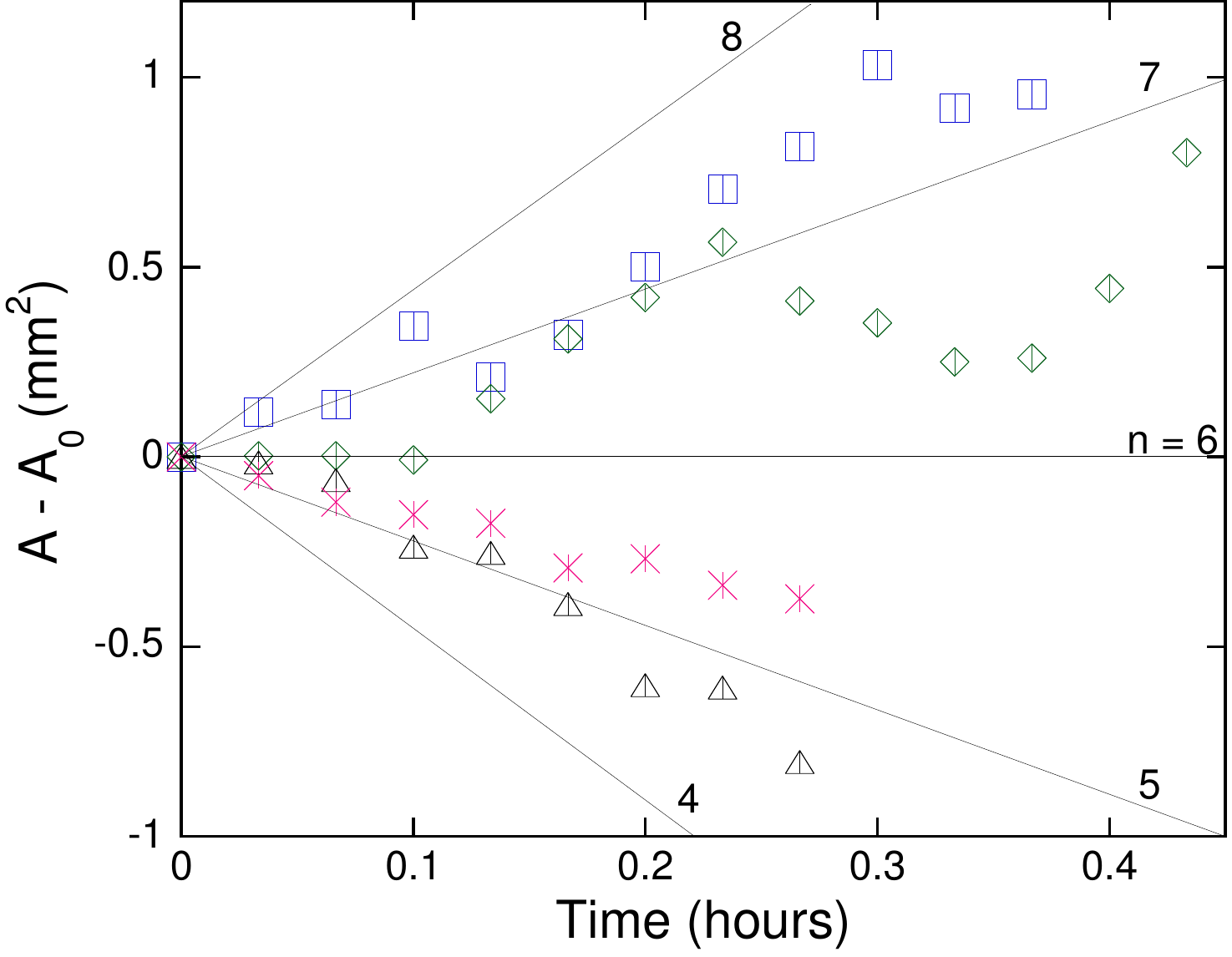}
\caption{(Color online)  Area versus time curves for four example six-sided bubbles.  The initial areas are subtracted off for ease of comparison.  Solid lines represent the average growth rates of $n$-sided bubbles according to Fig.~\ref{vnlawsurface}a.}
\label{sixsideavt}
\end{figure}

\begin{figure}
\includegraphics[width=3.00in]{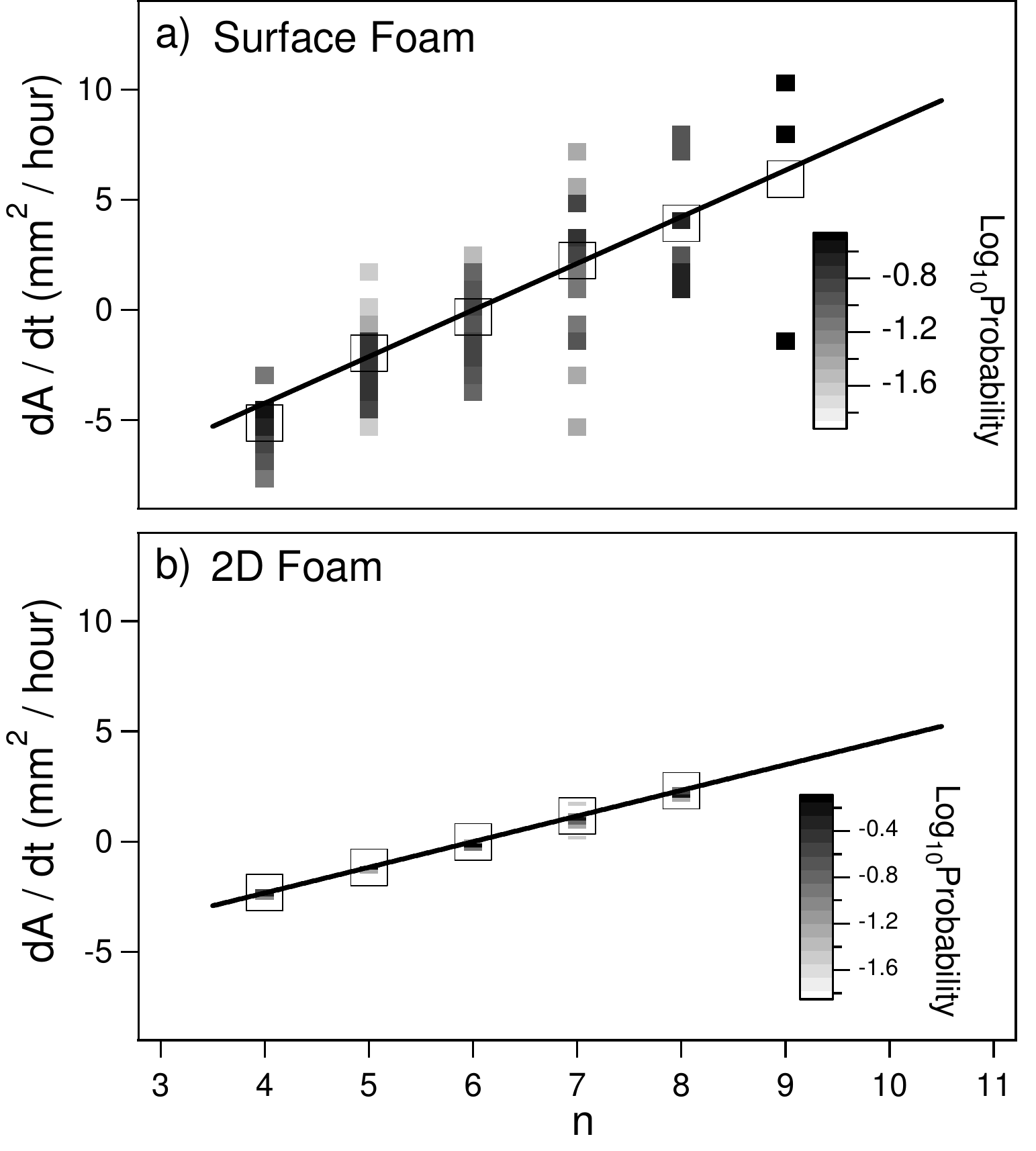}
\caption{Rate of change of area versus side number for all tracked bubbles for (a) surface foams and (b) the vertical two-dimensional foam from Ref.~\cite{RothDurian13}).  Greyscale shows the probability of an $n$-sided bubble having that coarsening rate.  Squares are the average rate of change of area for a given $n$.  The line is a fit to $\langle {\rm d}A_{n} / {\rm d}t \rangle = K (n-6)$, with $K = 2.2 \pm 0.1~{\rm mm^{2} / hour}$ for the surface foam and $K = 1.20 \pm 0.06~{\rm mm^{2} / hour}$ for the vertical two-dimensional foam from Ref.~\cite{RothDurian13}.}
\label{vnlawsurface}
\end{figure}

All measurements discussed to this point have involved individual static photographs and have not considered how the bubbles change over time.  An initial clear signal that the dynamics are different in the case of the surface foam is the creation of bubbles.  This type of T2 process is not possible in two-dimensional foams, but in our surface foam we do observe the creation of bubbles.  This is rare, and happens when bulk bubbles move to the surface.  The creation of surface bubbles by rearrangement occurred at an insignificant rate compared to the disappearance of bubbles by coarsening.

We are also able to track the change in individual bubbles over time.  In a sequence of 30 images, we measure the area of a bubble at each time, and fit to a line to determine ${\rm d}A/{\rm d}t$.  Only bubbles that did not change $n$ during this window were considered, so topological changes were not an issue.  For area versus time curves that were not linear, each linear region was considered separately.  In this way we can measure the coarsening rate of a large number of bubbles.  In a two-dimensional foam, the coarsening rate of an individual bubble depends only on the number of sides of that bubble, a surprising result known as von~Neumann's law, Eq.~(\ref{vn}).  In the case of our surface foam, we are only seeing the individual faces of larger three-dimensional bubbles, which can exchange gas through diffusion not just with the bubbles we can see, but others in the bulk.  The exact equation for the growth rate of a three-dimensional boundary bubble is shown in Eq.~(\ref{mpsbound}).  Additionally, the growth or shrinkage of bubble volume does not necessarily correspond to the area change of a single film.  The von~Neumann argument for two-dimensional foams thus cannot be applied and the coarsening rate of the two-dimensional surface bubbles will not be expected to depend only on the number of sides.  We expect that in three dimensions, larger bubbles and bubbles with more faces will be more likely to grow, so there should be some correlation between number of sides and area and coarsening rate.

In Fig.~\ref{sixsideavt} we plot area versus time curves for four example six sided bubbles.  In two dimensions, all six sided bubbles are stationary and neither grow nor shrink.  By contrast, the areas of the six-sided surface bubbles in Fig.~\ref{sixsideavt} all change with time.  Some grow, and some shrink, at a wide variety of coarsening rates -- occasionally exceeding the average growth rates for 5- and 7-sided bubbles.  We also note that the displayed area versus time curves are not linear, even though there are no topological changes during this time window.

The coarsening rate, ${\rm d}A/{\rm d}t$, for individual bubbles is plotted against number of sides in Fig.~\ref{vnlawsurface}a.  The density of points, indicated by greyscale, displays a large scatter of coarsening rates for bubbles with the same number of sides, and a large overlap of bubbles with different numbers of sides having the same coarsening rates.  This is very different from coarsening in a two-dimensional foam, shown in Fig.~\ref{vnlawsurface}b, where growth rates are tightly clustered around the average.
  
For the case of the surface foam, we expect the coarsening to be very complicated, with gas diffusion possible between surface bubbles and bubbles in the bulk, as well as a limited correlation between the change in bubble volume and the change in area of a single face on the surface.  Indeed, we see a wide range of growth rates in the case of the surface foam.  However, despite this wide variation for individual bubbles, we see that on average bubbles with more sides grow faster.  Remarkably, the average coarsening rate of $n$-sided bubbles can be well fit to a $K(n-6)$ proportionality like von~Neumann's law; this give $K = 2.2 \pm 0.1~{\rm mm^{2} / hour}$.  On average, von Neumann's law appears to hold, to within a margin of error that is small compared to the width of the distribution of rates found around the average.  While we thus detect no deviation from von~Neumann's law for the average behavior, except perhaps for $n=3$, it is possible that a deviation could be found by further experiments with better statistics.

\section{Conclusion}

We measured distributions and dynamics of the two-dimensional surface of a three-dimensional foam.  A total-internal reflection technique involving submerging the apparatus in water allowed us to cleanly image the surface, and image analysis allowed us to process a large quantity of data to build good statistics.

Some measurements were very similar to the case for an ordinary two-dimensional foam.  The number of sides distribution for the surface foam was slightly broader than for the two-dimensional case, but other topological measurements, such as $F(n)$ and $m(n)$, were nearly indistinguishable.  Likewise, the size distributions were very close in both cases.  There was a slight difference in the area distribution for large $A / \langle A \rangle$, but the perimeter distribution was the same in both cases.  The measurement of the shape parameter circularity was also the same for the surface foam and the two-dimensional foam.

The distribution of another shape parameter, elongation, was noticeably different in the surface foam.  Unlike the two-dimensional foam, for the surface foam there was a tail of large elongation bubbles.  This resulted in the average elongation being different for the two cases, although the distributions were peaked near the same value.

The greatest difference between the surface foam and the two-dimensional foam was in the dynamics.  Unlike the two-dimensional foam, which obeys von~Neumann's law, the surface foam had individual bubbles that coarsened at a wide variety of rates.  Additionally, we observed the creation of bubbles, a topological change that is not possible for two-dimensional foams.  Despite the spread in growth rates for the surface foam, von~Neumann's law appears to hold on average, to within a margin of error that is smaller than the widths of the growth rate distributions.  This surprising result remains to be explained, perhaps based on Eq.~(\ref{mpsbound}) and suitable assumptions about bubble sizes, shapes, and size-topology correlations.

\section{Acknowledgements}

This work was supported by NASA Microgravity Fluid Physics Grant NNX07AP20G (AER \& DJD) and NSF/DMR05-47230 (BGC).

\bibliography{CoarseningRefs}

\begin{thebibliography}{58}
\expandafter\ifx\csname natexlab\endcsname\relax\def\natexlab#1{#1}\fi
\expandafter\ifx\csname bibnamefont\endcsname\relax
  \def\bibnamefont#1{#1}\fi
\expandafter\ifx\csname bibfnamefont\endcsname\relax
  \def\bibfnamefont#1{#1}\fi
\expandafter\ifx\csname citenamefont\endcsname\relax
  \def\citenamefont#1{#1}\fi
\expandafter\ifx\csname url\endcsname\relax
  \def\url#1{\texttt{#1}}\fi
\expandafter\ifx\csname urlprefix\endcsname\relax\def\urlprefix{URL }\fi
\providecommand{\bibinfo}[2]{#2}
\providecommand{\eprint}[2][]{\url{#2}}

\bibitem[{\citenamefont{Weaire and Hutzler}(1999)}]{WeaireHutzlerBook}
\bibinfo{author}{\bibfnamefont{D.}~\bibnamefont{Weaire}} \bibnamefont{and}
  \bibinfo{author}{\bibfnamefont{S.}~\bibnamefont{Hutzler}},
  \emph{\bibinfo{title}{The Physics of Foams}} (\bibinfo{publisher}{Oxford
  University Press}, \bibinfo{address}{New York, NY}, \bibinfo{year}{1999}).

\bibitem[{\citenamefont{Matzke}(1946)}]{Matzke}
\bibinfo{author}{\bibfnamefont{E.~B.} \bibnamefont{Matzke}},
  \bibinfo{journal}{Am. J. Botany} \textbf{\bibinfo{volume}{33}},
  \bibinfo{pages}{58} (\bibinfo{year}{1946}).

\bibitem[{\citenamefont{Gonatas et~al.}(1995)\citenamefont{Gonatas, Leigh,
  Yodh, Glazier, and Prause}}]{Gonatusetal95}
\bibinfo{author}{\bibfnamefont{C.~P.} \bibnamefont{Gonatas}},
  \bibinfo{author}{\bibfnamefont{J.~S.} \bibnamefont{Leigh}},
  \bibinfo{author}{\bibfnamefont{A.~G.} \bibnamefont{Yodh}},
  \bibinfo{author}{\bibfnamefont{J.~A.} \bibnamefont{Glazier}},
  \bibnamefont{and} \bibinfo{author}{\bibfnamefont{B.}~\bibnamefont{Prause}},
  \bibinfo{journal}{Phys. Rev. Lett.} \textbf{\bibinfo{volume}{75}},
  \bibinfo{pages}{573} (\bibinfo{year}{1995}).

\bibitem[{\citenamefont{Monnereau and Vignes-Adler}(1998)}]{Adler98}
\bibinfo{author}{\bibfnamefont{C.}~\bibnamefont{Monnereau}} \bibnamefont{and}
  \bibinfo{author}{\bibfnamefont{M.}~\bibnamefont{Vignes-Adler}},
  \bibinfo{journal}{Phys. Rev. Lett.} \textbf{\bibinfo{volume}{80}},
  \bibinfo{pages}{5228} (\bibinfo{year}{1998}).

\bibitem[{\citenamefont{Fetterman et~al.}(2000)\citenamefont{Fetterman, Tan,
  Ying, Stack, Marks, Feller, Cull, Sullivan, Munson, Thoroddsen
  et~al.}}]{Fetterman00}
\bibinfo{author}{\bibfnamefont{M.~R.} \bibnamefont{Fetterman}},
  \bibinfo{author}{\bibfnamefont{E.}~\bibnamefont{Tan}},
  \bibinfo{author}{\bibfnamefont{L.}~\bibnamefont{Ying}},
  \bibinfo{author}{\bibfnamefont{R.~A.} \bibnamefont{Stack}},
  \bibinfo{author}{\bibfnamefont{D.~L.} \bibnamefont{Marks}},
  \bibinfo{author}{\bibfnamefont{S.}~\bibnamefont{Feller}},
  \bibinfo{author}{\bibfnamefont{E.}~\bibnamefont{Cull}},
  \bibinfo{author}{\bibfnamefont{J.~M.} \bibnamefont{Sullivan}},
  \bibinfo{author}{\bibfnamefont{J.}~\bibnamefont{Munson},
  \bibfnamefont{D.~C.}}, \bibinfo{author}{\bibfnamefont{S.~T.}
  \bibnamefont{Thoroddsen}}, \bibnamefont{et~al.}, \bibinfo{journal}{Optics
  Express} \textbf{\bibinfo{volume}{7}} (\bibinfo{year}{2000}).

\bibitem[{\citenamefont{Lambert et~al.}(2005)\citenamefont{Lambert, Cantat,
  Delannay, Renault, Graner, Glazier, Veretennikov, and
  Cloetens}}]{GlazierGraner05}
\bibinfo{author}{\bibfnamefont{J.}~\bibnamefont{Lambert}},
  \bibinfo{author}{\bibfnamefont{I.}~\bibnamefont{Cantat}},
  \bibinfo{author}{\bibfnamefont{R.}~\bibnamefont{Delannay}},
  \bibinfo{author}{\bibfnamefont{A.}~\bibnamefont{Renault}},
  \bibinfo{author}{\bibfnamefont{F.}~\bibnamefont{Graner}},
  \bibinfo{author}{\bibfnamefont{J.~A.} \bibnamefont{Glazier}},
  \bibinfo{author}{\bibfnamefont{I.}~\bibnamefont{Veretennikov}},
  \bibnamefont{and} \bibinfo{author}{\bibfnamefont{P.}~\bibnamefont{Cloetens}},
  \bibinfo{journal}{Coll. and Surf. A} \textbf{\bibinfo{volume}{263}},
  \bibinfo{pages}{295} (\bibinfo{year}{2005}).

\bibitem[{\citenamefont{Lambert et~al.}(2010)\citenamefont{Lambert, Mokso,
  Cantat, Cloetens, Glazier, Graner, and Delannay}}]{GlazierGraner10}
\bibinfo{author}{\bibfnamefont{J.}~\bibnamefont{Lambert}},
  \bibinfo{author}{\bibfnamefont{R.}~\bibnamefont{Mokso}},
  \bibinfo{author}{\bibfnamefont{I.}~\bibnamefont{Cantat}},
  \bibinfo{author}{\bibfnamefont{P.}~\bibnamefont{Cloetens}},
  \bibinfo{author}{\bibfnamefont{J.~A.} \bibnamefont{Glazier}},
  \bibinfo{author}{\bibfnamefont{F.}~\bibnamefont{Graner}}, \bibnamefont{and}
  \bibinfo{author}{\bibfnamefont{R.}~\bibnamefont{Delannay}},
  \bibinfo{journal}{Phys. Rev. Lett.} \textbf{\bibinfo{volume}{104}},
  \bibinfo{pages}{248304} (\bibinfo{year}{2010}).

\bibitem[{\citenamefont{Meagher et~al.}(2011)\citenamefont{Meagher, Mukherjee,
  Weaire, Hutzler, Banhart, and Garcia-Moreno}}]{Meagher11}
\bibinfo{author}{\bibfnamefont{A.~J.} \bibnamefont{Meagher}},
  \bibinfo{author}{\bibfnamefont{M.}~\bibnamefont{Mukherjee}},
  \bibinfo{author}{\bibfnamefont{D.}~\bibnamefont{Weaire}},
  \bibinfo{author}{\bibfnamefont{S.}~\bibnamefont{Hutzler}},
  \bibinfo{author}{\bibfnamefont{J.}~\bibnamefont{Banhart}}, \bibnamefont{and}
  \bibinfo{author}{\bibfnamefont{F.}~\bibnamefont{Garcia-Moreno}},
  \bibinfo{journal}{Soft Matter} \textbf{\bibinfo{volume}{7}},
  \bibinfo{pages}{9881} (\bibinfo{year}{2011}).

\bibitem[{\citenamefont{Glazier and Weaire}(1992)}]{GlazierWeaire92}
\bibinfo{author}{\bibfnamefont{J.}~\bibnamefont{Glazier}} \bibnamefont{and}
  \bibinfo{author}{\bibfnamefont{D.}~\bibnamefont{Weaire}},
  \bibinfo{journal}{J. Phys.: Condens. Matter} \textbf{\bibinfo{volume}{4}},
  \bibinfo{pages}{1867} (\bibinfo{year}{1992}).

\bibitem[{\citenamefont{Stavans}(1993{\natexlab{a}})}]{Stavans93}
\bibinfo{author}{\bibfnamefont{J.}~\bibnamefont{Stavans}},
  \bibinfo{journal}{Rep. Prog. Phys.} \textbf{\bibinfo{volume}{56}},
  \bibinfo{pages}{733} (\bibinfo{year}{1993}{\natexlab{a}}).

\bibitem[{\citenamefont{Durian et~al.}(1991{\natexlab{a}})\citenamefont{Durian,
  Weitz, and Pine}}]{DurianWeitzPine91a}
\bibinfo{author}{\bibfnamefont{D.~J.} \bibnamefont{Durian}},
  \bibinfo{author}{\bibfnamefont{D.~A.} \bibnamefont{Weitz}}, \bibnamefont{and}
  \bibinfo{author}{\bibfnamefont{D.~J.} \bibnamefont{Pine}},
  \bibinfo{journal}{Science} \textbf{\bibinfo{volume}{252}},
  \bibinfo{pages}{686} (\bibinfo{year}{1991}{\natexlab{a}}).

\bibitem[{\citenamefont{Durian et~al.}(1991{\natexlab{b}})\citenamefont{Durian,
  Weitz, and Pine}}]{DurianWeitzPine91b}
\bibinfo{author}{\bibfnamefont{D.~J.} \bibnamefont{Durian}},
  \bibinfo{author}{\bibfnamefont{D.~A.} \bibnamefont{Weitz}}, \bibnamefont{and}
  \bibinfo{author}{\bibfnamefont{D.~J.} \bibnamefont{Pine}},
  \bibinfo{journal}{Phys. Rev. A} \textbf{\bibinfo{volume}{44}},
  \bibinfo{pages}{7902} (\bibinfo{year}{1991}{\natexlab{b}}).

\bibitem[{\citenamefont{MacPherson and
  Srolovitz}(2007)}]{MacPhersonSrolovitz2007}
\bibinfo{author}{\bibfnamefont{R.~D.} \bibnamefont{MacPherson}}
  \bibnamefont{and} \bibinfo{author}{\bibfnamefont{D.~J.}
  \bibnamefont{Srolovitz}}, \bibinfo{journal}{Nature}
  \textbf{\bibinfo{volume}{446}}, \bibinfo{pages}{1053} (\bibinfo{year}{2007}).

\bibitem[{\citenamefont{Lambert et~al.}(2007)\citenamefont{Lambert, Cantat,
  Delannay, Mokso, Cloetens, Glazier, and Graner}}]{GlazierGraner07}
\bibinfo{author}{\bibfnamefont{J.}~\bibnamefont{Lambert}},
  \bibinfo{author}{\bibfnamefont{I.}~\bibnamefont{Cantat}},
  \bibinfo{author}{\bibfnamefont{R.}~\bibnamefont{Delannay}},
  \bibinfo{author}{\bibfnamefont{R.}~\bibnamefont{Mokso}},
  \bibinfo{author}{\bibfnamefont{P.}~\bibnamefont{Cloetens}},
  \bibinfo{author}{\bibfnamefont{J.~A.} \bibnamefont{Glazier}},
  \bibnamefont{and} \bibinfo{author}{\bibfnamefont{F.}~\bibnamefont{Graner}},
  \bibinfo{journal}{Phys. Rev. Lett.} \textbf{\bibinfo{volume}{99}},
  \bibinfo{pages}{058304} (\bibinfo{year}{2007}).

\bibitem[{\citenamefont{Glazier et~al.}(1987)\citenamefont{Glazier, Gross, and
  Stavans}}]{GlazierGrossStavans87}
\bibinfo{author}{\bibfnamefont{J.~A.} \bibnamefont{Glazier}},
  \bibinfo{author}{\bibfnamefont{S.~P.} \bibnamefont{Gross}}, \bibnamefont{and}
  \bibinfo{author}{\bibfnamefont{J.}~\bibnamefont{Stavans}},
  \bibinfo{journal}{Phys. Rev. A} \textbf{\bibinfo{volume}{36}},
  \bibinfo{pages}{306} (\bibinfo{year}{1987}).

\bibitem[{\citenamefont{Glazier and Stavans}(1989)}]{GlazierStavans89}
\bibinfo{author}{\bibfnamefont{J.~A.} \bibnamefont{Glazier}} \bibnamefont{and}
  \bibinfo{author}{\bibfnamefont{J.}~\bibnamefont{Stavans}},
  \bibinfo{journal}{Phys. Rev. A} \textbf{\bibinfo{volume}{40}},
  \bibinfo{pages}{7398} (\bibinfo{year}{1989}).

\bibitem[{\citenamefont{Stavans and Glazier}(1989)}]{StavansGlazier89}
\bibinfo{author}{\bibfnamefont{J.}~\bibnamefont{Stavans}} \bibnamefont{and}
  \bibinfo{author}{\bibfnamefont{J.~A.} \bibnamefont{Glazier}},
  \bibinfo{journal}{Phys. Rev. Lett.} \textbf{\bibinfo{volume}{62}},
  \bibinfo{pages}{1318} (\bibinfo{year}{1989}).

\bibitem[{\citenamefont{Stavans}(1990)}]{Stavans90}
\bibinfo{author}{\bibfnamefont{J.}~\bibnamefont{Stavans}},
  \bibinfo{journal}{Phys. Rev. A} \textbf{\bibinfo{volume}{42}},
  \bibinfo{pages}{5049} (\bibinfo{year}{1990}).

\bibitem[{\citenamefont{Stavans}(1993{\natexlab{b}})}]{Stavans93sf}
\bibinfo{author}{\bibfnamefont{J.}~\bibnamefont{Stavans}},
  \bibinfo{journal}{Physica A} \textbf{\bibinfo{volume}{194}},
  \bibinfo{pages}{307} (\bibinfo{year}{1993}{\natexlab{b}}).

\bibitem[{\citenamefont{de~Icaza et~al.}(1994)\citenamefont{de~Icaza,
  Jim$\acute{{\rm e}}$nez-Ceniceros, and Casta$\tilde{{\rm n}}$o}}]{Icaza94}
\bibinfo{author}{\bibfnamefont{M.}~\bibnamefont{de~Icaza}},
  \bibinfo{author}{\bibfnamefont{A.}~\bibnamefont{Jim$\acute{{\rm
  e}}$nez-Ceniceros}}, \bibnamefont{and} \bibinfo{author}{\bibfnamefont{V.~M.}
  \bibnamefont{Casta$\tilde{{\rm n}}$o}}, \bibinfo{journal}{J. Appl. Phys.}
  \textbf{\bibinfo{volume}{76}}, \bibinfo{pages}{7317} (\bibinfo{year}{1994}).

\bibitem[{\citenamefont{Roth et~al.}(2013)\citenamefont{Roth, Jones, and
  Durian}}]{RothDurian13}
\bibinfo{author}{\bibfnamefont{A.~E.} \bibnamefont{Roth}},
  \bibinfo{author}{\bibfnamefont{C.~D.} \bibnamefont{Jones}}, \bibnamefont{and}
  \bibinfo{author}{\bibfnamefont{D.~J.} \bibnamefont{Durian}},
  \bibinfo{journal}{Phys. Rev. E} \textbf{\bibinfo{volume}{87}},
  \bibinfo{pages}{042304} (\bibinfo{year}{2013}).

\bibitem[{\citenamefont{Krichevsky and Stavans}(1992)}]{StavansKrichevsky92}
\bibinfo{author}{\bibfnamefont{O.}~\bibnamefont{Krichevsky}} \bibnamefont{and}
  \bibinfo{author}{\bibfnamefont{J.}~\bibnamefont{Stavans}},
  \bibinfo{journal}{Phys. Rev. B} \textbf{\bibinfo{volume}{46}},
  \bibinfo{pages}{10579} (\bibinfo{year}{1992}).

\bibitem[{\citenamefont{Rosa and Fortes}(1999)}]{RosaFortes99}
\bibinfo{author}{\bibfnamefont{M.~E.} \bibnamefont{Rosa}} \bibnamefont{and}
  \bibinfo{author}{\bibfnamefont{M.~A.} \bibnamefont{Fortes}},
  \bibinfo{journal}{Philos. Mag. A} \textbf{\bibinfo{volume}{79}},
  \bibinfo{pages}{1871} (\bibinfo{year}{1999}).

\bibitem[{\citenamefont{Rosa et~al.}(2002)\citenamefont{Rosa, Afonso, and
  Fortes}}]{RosaFortes02}
\bibinfo{author}{\bibfnamefont{A.~E.} \bibnamefont{Rosa}},
  \bibinfo{author}{\bibfnamefont{L.}~\bibnamefont{Afonso}}, \bibnamefont{and}
  \bibinfo{author}{\bibfnamefont{M.~A.} \bibnamefont{Fortes}},
  \bibinfo{journal}{Philos. Mag. A} \textbf{\bibinfo{volume}{82}},
  \bibinfo{pages}{2953} (\bibinfo{year}{2002}).

\bibitem[{\citenamefont{Stine et~al.}(1990)\citenamefont{Stine, Rauseo, Moore,
  Wise, and Knober}}]{SteinMoore90}
\bibinfo{author}{\bibfnamefont{K.~J.} \bibnamefont{Stine}},
  \bibinfo{author}{\bibfnamefont{S.~A.} \bibnamefont{Rauseo}},
  \bibinfo{author}{\bibfnamefont{B.~G.} \bibnamefont{Moore}},
  \bibinfo{author}{\bibfnamefont{J.~A.} \bibnamefont{Wise}}, \bibnamefont{and}
  \bibinfo{author}{\bibfnamefont{C.~M.} \bibnamefont{Knober}},
  \bibinfo{journal}{Phys. Rev. A} \textbf{\bibinfo{volume}{41}},
  \bibinfo{pages}{6884} (\bibinfo{year}{1990}).

\bibitem[{\citenamefont{Berge et~al.}(1990)\citenamefont{Berge, Simon, and
  Libchaber}}]{Bergeetal90}
\bibinfo{author}{\bibfnamefont{B.}~\bibnamefont{Berge}},
  \bibinfo{author}{\bibfnamefont{A.~J.} \bibnamefont{Simon}}, \bibnamefont{and}
  \bibinfo{author}{\bibfnamefont{A.}~\bibnamefont{Libchaber}},
  \bibinfo{journal}{Phys. Rev. A} \textbf{\bibinfo{volume}{41}},
  \bibinfo{pages}{6893} (\bibinfo{year}{1990}).

\bibitem[{\citenamefont{Kermode and Weaire}(1990)}]{KermodeWeaire90}
\bibinfo{author}{\bibfnamefont{J.~P.} \bibnamefont{Kermode}} \bibnamefont{and}
  \bibinfo{author}{\bibfnamefont{D.}~\bibnamefont{Weaire}},
  \bibinfo{journal}{Computer Phys. Comm.} \textbf{\bibinfo{volume}{60}},
  \bibinfo{pages}{75} (\bibinfo{year}{1990}).

\bibitem[{\citenamefont{Glazier et~al.}(1990)\citenamefont{Glazier, Anderson,
  and Grest}}]{GlazierAndersonGrest90}
\bibinfo{author}{\bibfnamefont{J.}~\bibnamefont{Glazier}},
  \bibinfo{author}{\bibfnamefont{M.}~\bibnamefont{Anderson}}, \bibnamefont{and}
  \bibinfo{author}{\bibfnamefont{G.}~\bibnamefont{Grest}},
  \bibinfo{journal}{Philos. Mag. B} \textbf{\bibinfo{volume}{62}},
  \bibinfo{pages}{615} (\bibinfo{year}{1990}).

\bibitem[{\citenamefont{Herdtle and Aref}(1992)}]{HerdtleAref92}
\bibinfo{author}{\bibfnamefont{T.}~\bibnamefont{Herdtle}} \bibnamefont{and}
  \bibinfo{author}{\bibfnamefont{H.}~\bibnamefont{Aref}}, \bibinfo{journal}{J.
  Fluid Mech.} \textbf{\bibinfo{volume}{241}}, \bibinfo{pages}{233}
  (\bibinfo{year}{1992}).

\bibitem[{\citenamefont{Segel et~al.}(1993)\citenamefont{Segel, Mukamel,
  Krichevsky, and Stavans}}]{Segeletal93}
\bibinfo{author}{\bibfnamefont{D.}~\bibnamefont{Segel}},
  \bibinfo{author}{\bibfnamefont{D.}~\bibnamefont{Mukamel}},
  \bibinfo{author}{\bibfnamefont{O.}~\bibnamefont{Krichevsky}},
  \bibnamefont{and} \bibinfo{author}{\bibfnamefont{J.}~\bibnamefont{Stavans}},
  \bibinfo{journal}{Phys. Rev. E} \textbf{\bibinfo{volume}{47}},
  \bibinfo{pages}{812} (\bibinfo{year}{1993}).

\bibitem[{\citenamefont{Neubert and Schreckenberg}(1997)}]{NeuSch97}
\bibinfo{author}{\bibfnamefont{L.}~\bibnamefont{Neubert}} \bibnamefont{and}
  \bibinfo{author}{\bibfnamefont{M.}~\bibnamefont{Schreckenberg}},
  \bibinfo{journal}{Physica A} \textbf{\bibinfo{volume}{240}},
  \bibinfo{pages}{491} (\bibinfo{year}{1997}).

\bibitem[{\citenamefont{Rutenberg and McCurdy}(2006)}]{Rutenberg05}
\bibinfo{author}{\bibfnamefont{A.~D.} \bibnamefont{Rutenberg}}
  \bibnamefont{and} \bibinfo{author}{\bibfnamefont{M.~B.}
  \bibnamefont{McCurdy}}, \bibinfo{journal}{Phys. Rev. E}
  \textbf{\bibinfo{volume}{73}}, \bibinfo{pages}{011403}
  (\bibinfo{year}{2006}).

\bibitem[{\citenamefont{Fortuna et~al.}(2012)\citenamefont{Fortuna, Thomas,
  de~Almeida, and Graner}}]{Graneretalarxiv}
\bibinfo{author}{\bibfnamefont{I.}~\bibnamefont{Fortuna}},
  \bibinfo{author}{\bibfnamefont{G.~L.} \bibnamefont{Thomas}},
  \bibinfo{author}{\bibfnamefont{R.~M.~C.} \bibnamefont{de~Almeida}},
  \bibnamefont{and} \bibinfo{author}{\bibfnamefont{F.}~\bibnamefont{Graner}},
  \bibinfo{journal}{Phys. Rev. Lett.} \textbf{\bibinfo{volume}{108}},
  \bibinfo{pages}{248301} (\bibinfo{year}{2012}).

\bibitem[{\citenamefont{von Neumann}(1952)}]{VonNeumann}
\bibinfo{author}{\bibfnamefont{J.}~\bibnamefont{von Neumann}}, in
  \emph{\bibinfo{booktitle}{Metal Interfaces}} (\bibinfo{publisher}{American
  Society for Metals}, \bibinfo{address}{Cleveland}, \bibinfo{year}{1952}), pp.
  \bibinfo{pages}{108--110}.

\bibitem[{\citenamefont{Mancini and Oguey}(2005{\natexlab{a}})}]{Mancini05a}
\bibinfo{author}{\bibfnamefont{M.}~\bibnamefont{Mancini}} \bibnamefont{and}
  \bibinfo{author}{\bibfnamefont{C.}~\bibnamefont{Oguey}},
  \bibinfo{journal}{Eur. Phys. J. E} \textbf{\bibinfo{volume}{17}},
  \bibinfo{pages}{119} (\bibinfo{year}{2005}{\natexlab{a}}).

\bibitem[{\citenamefont{Mancini and Oguey}(2005{\natexlab{b}})}]{Mancini05b}
\bibinfo{author}{\bibfnamefont{M.}~\bibnamefont{Mancini}} \bibnamefont{and}
  \bibinfo{author}{\bibfnamefont{C.}~\bibnamefont{Oguey}},
  \bibinfo{journal}{Coll. and Surf. A} \textbf{\bibinfo{volume}{263}},
  \bibinfo{pages}{33} (\bibinfo{year}{2005}{\natexlab{b}}).

\bibitem[{\citenamefont{Clark and Blackman}(1948)}]{ClarkBlackman48}
\bibinfo{author}{\bibfnamefont{N.~O.} \bibnamefont{Clark}} \bibnamefont{and}
  \bibinfo{author}{\bibfnamefont{M.}~\bibnamefont{Blackman}},
  \bibinfo{journal}{Transactions of the Faraday Society}
  \textbf{\bibinfo{volume}{44}}, \bibinfo{pages}{350} (\bibinfo{year}{1948}).

\bibitem[{\citenamefont{Chang et~al.}(1956)\citenamefont{Chang, Schoen, and
  Grove}}]{ChangSchoenGrove56}
\bibinfo{author}{\bibfnamefont{R.~C.} \bibnamefont{Chang}},
  \bibinfo{author}{\bibfnamefont{H.~M.} \bibnamefont{Schoen}},
  \bibnamefont{and} \bibinfo{author}{\bibfnamefont{C.~S.} \bibnamefont{Grove}},
  \bibinfo{journal}{Industrial and Engineering Chemistry}
  \textbf{\bibinfo{volume}{48}}, \bibinfo{pages}{2035} (\bibinfo{year}{1956}).

\bibitem[{\citenamefont{de~Vries}(1957)}]{deVries57}
\bibinfo{author}{\bibfnamefont{A.~J.} \bibnamefont{de~Vries}}, in
  \emph{\bibinfo{booktitle}{Foam Stability}}
  (\bibinfo{publisher}{Rubber-Stichting}, \bibinfo{address}{Delft},
  \bibinfo{year}{1957}).

\bibitem[{\citenamefont{Jashnani and Lemlich}(1974)}]{JashnaniLemlich74}
\bibinfo{author}{\bibfnamefont{I.~L.} \bibnamefont{Jashnani}} \bibnamefont{and}
  \bibinfo{author}{\bibfnamefont{R.}~\bibnamefont{Lemlich}},
  \bibinfo{journal}{Colloid and Interface Science}
  \textbf{\bibinfo{volume}{46}}, \bibinfo{pages}{13} (\bibinfo{year}{1974}).

\bibitem[{\citenamefont{Feitosa et~al.}(2006)\citenamefont{Feitosa, Halt,
  Kamien, and Durian}}]{FeitosaDurian06}
\bibinfo{author}{\bibfnamefont{K.}~\bibnamefont{Feitosa}},
  \bibinfo{author}{\bibfnamefont{O.~L.} \bibnamefont{Halt}},
  \bibinfo{author}{\bibfnamefont{R.~D.} \bibnamefont{Kamien}},
  \bibnamefont{and} \bibinfo{author}{\bibfnamefont{D.~J.}
  \bibnamefont{Durian}}, \bibinfo{journal}{Europhys. Lett.}
  \textbf{\bibinfo{volume}{76}}, \bibinfo{pages}{683} (\bibinfo{year}{2006}).

\bibitem[{\citenamefont{Gardiner et~al.}(2000)\citenamefont{Gardiner,
  Dlugogorski, and Jameson}}]{Jameson99}
\bibinfo{author}{\bibfnamefont{B.~S.} \bibnamefont{Gardiner}},
  \bibinfo{author}{\bibfnamefont{B.~Z.} \bibnamefont{Dlugogorski}},
  \bibnamefont{and} \bibinfo{author}{\bibfnamefont{G.~J.}
  \bibnamefont{Jameson}}, \bibinfo{journal}{Philos. Mag. A}
  \textbf{\bibinfo{volume}{80}}, \bibinfo{pages}{981} (\bibinfo{year}{2000}).

\bibitem[{\citenamefont{de~Vries}(1972)}]{deVries72}
\bibinfo{author}{\bibfnamefont{A.~J.} \bibnamefont{de~Vries}}, in
  \emph{\bibinfo{booktitle}{Adsorptive Bubble Separation Techniques}}, edited
  by \bibinfo{editor}{\bibfnamefont{R.}~\bibnamefont{Lemlich}}
  (\bibinfo{publisher}{Academic Press}, \bibinfo{address}{New York},
  \bibinfo{year}{1972}), chap.~\bibinfo{chapter}{2}.

\bibitem[{\citenamefont{Cheng and Lemlich}(1983)}]{ChengLemlich}
\bibinfo{author}{\bibfnamefont{H.~C.} \bibnamefont{Cheng}} \bibnamefont{and}
  \bibinfo{author}{\bibfnamefont{R.}~\bibnamefont{Lemlich}},
  \bibinfo{journal}{Industrial and Engineering Chemistry Fundamentals}
  \textbf{\bibinfo{volume}{22}}, \bibinfo{pages}{105} (\bibinfo{year}{1983}).

\bibitem[{\citenamefont{Wang and Neethling}(2006)}]{WangNeethling06}
\bibinfo{author}{\bibfnamefont{Y.}~\bibnamefont{Wang}} \bibnamefont{and}
  \bibinfo{author}{\bibfnamefont{S.~J.} \bibnamefont{Neethling}},
  \bibinfo{journal}{Minerals Engineering} \textbf{\bibinfo{volume}{19}},
  \bibinfo{pages}{1069} (\bibinfo{year}{2006}).

\bibitem[{\citenamefont{Wang and Neethling}(2009)}]{WangNeethling09}
\bibinfo{author}{\bibfnamefont{Y.}~\bibnamefont{Wang}} \bibnamefont{and}
  \bibinfo{author}{\bibfnamefont{S.~J.} \bibnamefont{Neethling}},
  \bibinfo{journal}{Colloids and Surfaces A} \textbf{\bibinfo{volume}{339}},
  \bibinfo{pages}{73} (\bibinfo{year}{2009}).

\bibitem[{\citenamefont{Mason et~al.}()\citenamefont{Mason, Lazar, MacPherson,
  and Srolovitz}}]{LazarSrolovitz}
\bibinfo{author}{\bibfnamefont{J.~K.} \bibnamefont{Mason}},
  \bibinfo{author}{\bibfnamefont{E.~A.} \bibnamefont{Lazar}},
  \bibinfo{author}{\bibfnamefont{R.~D.} \bibnamefont{MacPherson}},
  \bibnamefont{and} \bibinfo{author}{\bibfnamefont{D.~J.}
  \bibnamefont{Srolovitz}}, \bibinfo{note}{``Characterizing the Steady State
  Microstructure of Two- and Three-Dimensional Grain Growth'' (preprint,
  2013)}.

\bibitem[{\citenamefont{Fortes et~al.}(2002)\citenamefont{Fortes, Rosa, and
  Afonso}}]{RosaFortes02a}
\bibinfo{author}{\bibfnamefont{M.~A.} \bibnamefont{Fortes}},
  \bibinfo{author}{\bibfnamefont{M.~E.} \bibnamefont{Rosa}}, \bibnamefont{and}
  \bibinfo{author}{\bibfnamefont{L.}~\bibnamefont{Afonso}},
  \bibinfo{journal}{Phil. Mag. A} \textbf{\bibinfo{volume}{82}},
  \bibinfo{pages}{527} (\bibinfo{year}{2002}).

\bibitem[{\citenamefont{Vasconcelos et~al.}(2003)\citenamefont{Vasconcelos,
  Cantat, and Glazier}}]{VCG03}
\bibinfo{author}{\bibfnamefont{I.~F.} \bibnamefont{Vasconcelos}},
  \bibinfo{author}{\bibfnamefont{I.}~\bibnamefont{Cantat}}, \bibnamefont{and}
  \bibinfo{author}{\bibfnamefont{J.~A.} \bibnamefont{Glazier}},
  \bibinfo{journal}{Journal of Computational Physics}
  \textbf{\bibinfo{volume}{192}}, \bibinfo{pages}{1} (\bibinfo{year}{2003}).

\bibitem[{\citenamefont{Chiu}(1995)}]{ChiuReview}
\bibinfo{author}{\bibfnamefont{S.~N.} \bibnamefont{Chiu}},
  \bibinfo{journal}{Materials Characterization} \textbf{\bibinfo{volume}{34}},
  \bibinfo{pages}{149} (\bibinfo{year}{1995}).

\bibitem[{\citenamefont{Lewis}(1928)}]{Lewis1928}
\bibinfo{author}{\bibfnamefont{F.~T.} \bibnamefont{Lewis}},
  \bibinfo{journal}{Anat. Rec.} \textbf{\bibinfo{volume}{38}},
  \bibinfo{pages}{341} (\bibinfo{year}{1928}).

\bibitem[{\citenamefont{Lewis}(1930)}]{Lewis1930}
\bibinfo{author}{\bibfnamefont{F.~T.} \bibnamefont{Lewis}},
  \bibinfo{journal}{Anat. Rec.} \textbf{\bibinfo{volume}{47}},
  \bibinfo{pages}{59} (\bibinfo{year}{1930}).

\bibitem[{\citenamefont{Rivier and Lissowski}(1982)}]{RivierLissowski82}
\bibinfo{author}{\bibfnamefont{N.}~\bibnamefont{Rivier}} \bibnamefont{and}
  \bibinfo{author}{\bibfnamefont{A.}~\bibnamefont{Lissowski}},
  \bibinfo{journal}{J. Phys. A: Math. Gen.} \textbf{\bibinfo{volume}{15}},
  \bibinfo{pages}{143} (\bibinfo{year}{1982}).

\bibitem[{\citenamefont{Rivier}(1985)}]{Rivier85}
\bibinfo{author}{\bibfnamefont{N.}~\bibnamefont{Rivier}},
  \bibinfo{journal}{Phil. Mag. B} \textbf{\bibinfo{volume}{52}},
  \bibinfo{pages}{795} (\bibinfo{year}{1985}).

\bibitem[{\citenamefont{Szeto and Tam}(1995)}]{SzetoTam95}
\bibinfo{author}{\bibfnamefont{K.~Y.} \bibnamefont{Szeto}} \bibnamefont{and}
  \bibinfo{author}{\bibfnamefont{W.~Y.} \bibnamefont{Tam}},
  \bibinfo{journal}{Physica A} \textbf{\bibinfo{volume}{221}},
  \bibinfo{pages}{256} (\bibinfo{year}{1995}).

\bibitem[{\citenamefont{Saraiva et~al.}(2009)\citenamefont{Saraiva, Pina,
  Bandeira, and Antunes}}]{Saraivaetal09}
\bibinfo{author}{\bibfnamefont{J.}~\bibnamefont{Saraiva}},
  \bibinfo{author}{\bibfnamefont{P.}~\bibnamefont{Pina}},
  \bibinfo{author}{\bibfnamefont{L.}~\bibnamefont{Bandeira}}, \bibnamefont{and}
  \bibinfo{author}{\bibfnamefont{J.}~\bibnamefont{Antunes}},
  \bibinfo{journal}{Philos. Mag. Lett.} \textbf{\bibinfo{volume}{89}},
  \bibinfo{pages}{185} (\bibinfo{year}{2009}).

\bibitem[{\citenamefont{Durand et~al.}(2011)\citenamefont{Durand,
  K$\ddot{\textrm{a}}$fer, Quilliet, Cox, Talebi, and Graner}}]{Graneretal11}
\bibinfo{author}{\bibfnamefont{M.}~\bibnamefont{Durand}},
  \bibinfo{author}{\bibfnamefont{J.}~\bibnamefont{K$\ddot{\textrm{a}}$fer}},
  \bibinfo{author}{\bibfnamefont{C.}~\bibnamefont{Quilliet}},
  \bibinfo{author}{\bibfnamefont{S.}~\bibnamefont{Cox}},
  \bibinfo{author}{\bibfnamefont{S.~A.} \bibnamefont{Talebi}},
  \bibnamefont{and} \bibinfo{author}{\bibfnamefont{F.}~\bibnamefont{Graner}},
  \bibinfo{journal}{Phys. Rev. Lett.} \textbf{\bibinfo{volume}{107}},
  \bibinfo{pages}{168304} (\bibinfo{year}{2011}).

\bibitem[{\citenamefont{Lambert et~al.}(2012)\citenamefont{Lambert, Graner,
  Delannay, and Cantat}}]{Lambert2012}
\bibinfo{author}{\bibfnamefont{J.}~\bibnamefont{Lambert}},
  \bibinfo{author}{\bibfnamefont{F.}~\bibnamefont{Graner}},
  \bibinfo{author}{\bibfnamefont{R.}~\bibnamefont{Delannay}}, \bibnamefont{and}
  \bibinfo{author}{\bibfnamefont{I.}~\bibnamefont{Cantat}},
  \bibinfo{journal}{Europhys. Lett.} \textbf{\bibinfo{volume}{99}},
  \bibinfo{pages}{48003} (\bibinfo{year}{2012}).

\end{thebibliography}

\end{document}